\documentclass[letterpaper, journal]{IEEEtran}

\newtheorem{corollary}{Corollary}
\newtheorem{definition}{Definition}
\newtheorem{example}{Example}
\newtheorem{lemma}{Lemma}
\newtheorem{theorem}{Theorem}

\usepackage{amsmath, amsfonts}
\usepackage{amssymb}

\usepackage[noadjust]{cite}

\usepackage{geometry}
\usepackage[pdftex]{graphicx}

\usepackage{tikz}
\usetikzlibrary{calc, intersections, positioning}
\tikzset{> = latex}

\newcommand{\figuresname}{Figs.}
\newcommand{\markovlink}{\leftrightarrow} % Markov 链的一环

\newcommand{\entropy}[1]{H ( #1 )}

\newcommand{\mean}[1]{E [ #1 ]}
\newcommand{\minfo}[1]{I ( #1 )} % mutual information
\newcommand{\probability}[1]{\mathrm{Pr} [ #1 ]}

\newcommand{\cointegers}[2]{[ #1 : #2 )} % 左闭右开区间中的整数组成的集合
\newcommand{\cointerval}[2]{[ #1 , #2 )} % 左闭右开区间
\newcommand{\conumbers}[1]{[ #1 )}

\newcommand{\ocintegers}[2]{( #1 : #2 ]} % 左开右闭区间中的整数组成的集合
\newcommand{\ocnumbers}[1]{( #1 ]}
\newcommand{\oointerval}[2]{( #1 , #2 )} % 开区间
\newcommand{\reals}{\mathbb{R}}

\newcommand{\dnormal}[2]{\mathcal{N} ( #1 , #2 )} % normal distribution

\title{%
    A Rate-Distortion Analysis for Composite Sources Under Subsource-Dependent
    Fidelity Criteria%
}
\author{%
    Jiakun~Liu,~\IEEEmembership{Graduate~Student~Member,~IEEE,}
    H.~Vincent~Poor,~\IEEEmembership{Life~Fellow,~IEEE,}
    Iickho~Song,~\IEEEmembership{Fellow,~IEEE,}
    and~Wenyi~Zhang,~\IEEEmembership{Senior~Member,~IEEE}%
    \thanks{%
        Jiakun Liu and Wenyi Zhang are with the Department of Electronic
        Engineering and Information Science, University of Science and
        Technology of China, Hefei 230027, China (e-mail:
        liujk@mail.ustc.edu.cn; wenyizha@ustc.edu.cn).%
    }%
    \thanks{%
        H. Vincent Poor is with the Department of Electrical Engineering,
        Princeton University, Princeton, NJ 08544, USA (e-mail:
        poor@princeton.edu).%
    }%
    \thanks{%
        Iickho Song is with the School of Electrical Engineering, Korea
        Advanced Institute of Science and Technology, 291 Daehag Ro, Yuseong
        Gu, Daejeon 34141, Republic of Korea (e-mail: i.song@ieee.org).%
    }%
}

\begin{document}
    \maketitle
    \pagestyle{empty}
    \thispagestyle{empty}

    \begin{abstract}
        A composite source, consisting of multiple subsources and a memoryless
        switch, outputs one symbol at a time from the subsource selected by the
        switch.
        If some data should be encoded more accurately than other data from an
        information source, the composite source model is suitable because in
        this model different distortion constraints can be put on the
        subsources.
        In this context, we propose subsource-dependent fidelity criteria for
        composite sources and use them to formulate a rate-distortion problem.
        We solve the problem and obtain a single-letter expression for the
        rate-distortion function.
        Further rate-distortion analysis characterizes the performance of
        classify-then-compress (CTC) coding, which is frequently used in
        practice when subsource-dependent fidelity criteria are considered.
        Our analysis shows that CTC coding generally has performance loss
        relative to optimal coding, even if the classification is perfect.
        We also identify the cause of the performance loss, that is, class
        labels have to be reproduced in CTC coding.
        Last but not least, we show that the performance loss is negligible for
        asymptotically small distortion if CTC coding is appropriately designed
        and some mild conditions are satisfied.
    \end{abstract}


    % cspell: words ROIs, subsource, subsources

\section{Introduction}
\label{s:introduction}

\begin{figure}[!t]
    \centering \footnotesize
    % cspell: words sourceell, subsource

\begin{tikzpicture}[
    block/.style = {draw, minimum height = 4ex, minimum width = 5em}]

    % 子信源
    \matrix[row sep = 1ex] {
        \node[block] (source1) {Subsource~1}; \\
        % ``ss'' 表示 ``subsource''.
        \node[block] (source2) {Subsource~2}; \\
        \node (dots) {$\vdots$}; \\
        \node[block] (sourceell) {Subsource~$L$}; \\};

    % 用小圆表示的开关及其右边的线
    \node[draw, inner sep = 0, minimum size = 0.7ex, shape = circle] (root) at
    ($ 0.5*(source1) + 0.5*(sourceell) + (9em, 0) $) {};
    \node[right] (state) at ($(sourceell) + (9em, 0) $)
    {State $S \in \{ 1 , 2 , \cdots , L \}$};
    \draw[->] (state) -- (root);
    \draw (root) -- ++(3em, 0) node[align = left, right]
    {Symbol $X$ produced\\by Subsource $S$};

    % 连接子信源和开关的线
    \path[name path = hs1] (source1) -- ++(9em, 0);
    % ``hs'' 表示 ``horizontal segment''.
    \path[name path = ss1] (root) -- ++(120: 5em);
    % ``ss'' 表示 ``sloped segment''.
    \draw[name intersections = {of = hs1 and ss1}] (source1) --
    (intersection-1) -- ($ (root) + (120: 2em) $);
    \path[name path = hs2] (source2) -- ++(9em, 0);
    \path[name path = ss2] (root) -- ++(160: 5em);
    \draw[name intersections = {of = hs2 and ss2}] (source2) --
    (intersection-1) -- ($ (root) + (160: 2em) $);
    \draw ($ (root) + (200: 2em) $) -- (root);
    \path[name path = hsell] (sourceell) -- ++(9em, 0);
    \path[name path = ssell] (root) -- ++(240: 5em);
    \draw[name intersections = {of = hsell and ssell}] (sourceell) --
    (intersection-1) -- ($ (root) + (240: 2em) $);
\end{tikzpicture}
    \caption{A composite source with $L$ subsources.}
    \label{f:introduction.compsource}
\end{figure}
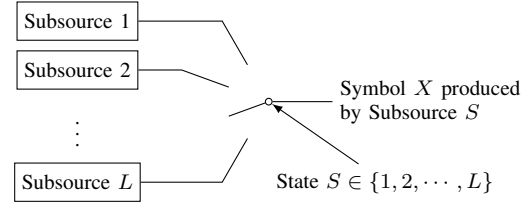

\IEEEPARstart{I}{t} is a common practice in rate-distortion (RD) theory
\cite{shannon1948, shannon1959, berger1971theory, cover2006, csiszar2011} to
model an information source as a stochastic process
$\{ X_{i} \}_{i = 1}^{\infty}$.
Each random variable $X_{i}$ is called a source symbol or just a symbol.
Symbols of many real-world sources can be naturally divided into a few classes.
For example, a pixel in a digital image may be in the foreground or background
\cite{szeliski2011}, and a sample in a piece of speech signal may belong to a
voiced phoneme, an unvoiced phoneme, or a period of silence \cite{rabiner2007}.
Symbols in different classes usually have different statistical characteristics
and are produced by different physical mechanisms.
Thus, it is not unreasonable to think of them as being produced by different
subsources, as done in the composite source model (see, for example,
\cite[Section~6.1]{berger1971theory}).
Besides the subsources, a composite source contains a conceptual switch, whose
position may vary with time.
As shown in \figurename~\ref{f:introduction.compsource}, the source outputs a
symbol produced by Subsource~$S$, which is selected by the switch.
We refer to $S$ as the state of the composite source.
A composite source is modeled as a stochastic process
$\{ ( S_{i} , X_{i} ) \}_{i = 1}^{\infty}$ of state-symbol pairs, where
$\{ S_{i} \}_{i = 1}^{\infty}$ take values in a common finite set, say
$\{ 1 , 2 , \cdots , L \}$.

In lossy compression of composite sources, we may wish to reproduce the symbols
from some subsources more accurately than the symbols from other subsources.
For example, an obscure foreground has a larger impact on the quality of an
image than an obscure background does.
Thus, the distortion constraint on the background can be loosened when we have
to compress the image into fewer bits.
Similarly, a sample of an unvoiced phoneme is more important than a sample of a
silence period, and therefore the fidelity of the former sample should be
guaranteed first in low-rate coding of speech.
This idea applies to general information sources and has been implemented in
some source coding techniques such as JPEG, where pixels in regions of interest
(ROIs) are coded with higher quality \cite{christopoulos2000}.

In this paper, we propose a new type of fidelity criteria, called
subsource-dependent fidelity criteria, that impose different distortion
constraints on the subsources.
We study an RD problem based on the criteria and obtain a single-letter
expression for the RD function.
The classical fidelity criterion can be viewed as a special case of the
proposed criteria.

Subsource-dependent fidelity criteria reflect the need to treat subsources of a
composite source differently, and a technique widely adopted to fulfill the
need is classify-then-compress (CTC) coding, i.e. dividing the source symbols
into a few classes and then compressing the classes separately.
See Section~\ref{u:introduction.works} for examples.
Here the idea behind CTC coding is that if most symbols in a class are produced
by the same subsource, then the class can be compressed according to distortion
constraints on the subsource.
In this paper, we fix an arbitrary classifier and design the coding of the
individual classes for good CTC coding performance.
The optimal performance is characterized by another RD function.
Comparison of the RD functions provides interesting insights, including:
\begin{enumerate}
    \item CTC coding in general has performance loss relative to RD-optimal
    coding, even if the classification is perfect;
    \item The cause of the performance loss is that class labels have to be
    reproduced in CTC coding, while separate compression of source symbols in
    different classes in CTC coding does not cause performance loss; and
    \item The performance loss of CTC coding is negligible for asymptotically
    small distortion if the coding is appropriately designed and some mild
    conditions are satisfied.
\end{enumerate}

A difficulty in the analysis of CTC coding is how to establish a mathematical
model for CTC coding, which we address by using variable-length codes.
A key tool in our proofs is a generalization of the conditional RD function
\cite{gray1972, gray1973} to subsource-dependent fidelity criteria, which may
be of independent interest.

This paper is organized as follows.
In the remaining part of this section we introduce related works and our
notation.
In Section~\ref{s:definitions}, we propose subsource-dependent fidelity
criteria and present other definitions that will be used in the following
sections.
In Section~\ref{s:results}, we present our main results, including coding
theorems and analysis on CTC coding.
Section~\ref{s:sketches} outlines the proofs of the coding theorems and
Section~\ref{s:conclusion} concludes the paper.
Detailed proofs are given in the appendices.

% cspell: words autoencoders, memoryless, quantizers, ROIs, subsource
% cspell: words subsources, uncountably, Wyner

\subsection{Related Works}
\label{u:introduction.works}

The first theoretical studies of the lossy compression problem were by Shannon
\cite{shannon1948, shannon1959}.
Later many variants of the problem were considered.
Many variants involve some side information, which is usually a correlated
source.
For example, if both the encoder and the decoder know the side information, the
problem becomes Gray's conditional RD problem \cite{gray1972, gray1973}.
If only the decoder knows the side information, the problem becomes the one
studied by Wyner and Ziv in \cite{wyner1976}.
If the side information is chosen by some adversary, the problem becomes
Berger's source coding game \cite[Section~6.1.2]{berger1971theory}
\cite{berger1971game}.
The side information can also influence distortion measures
\cite{linder2000, martinian2008}.
Each of the variants has a stochastic process, besides the information source,
that influences encoding of source symbols, decoding of source symbols, or
evaluation of distortion.
Our work considers another possibility that the stochastic process composed of
source states affects how distortion measures are applied to source symbols.
Another important variant of the lossy compression problem is the indirect RD
problem \cite[Section~3.5]{berger1971theory}
\cite{dobrushin1962, wolf1970, witsenhausen1980}, where the encoder observes
noisy versions of source symbols instead of source symbols themselves.
A source state in our work is similar to a source symbol in the indirect RD
problem in that neither of them can be observed directly.
Nevertheless, a source state is not a symbol, and need not be reproduced.

The composite source model was studied as early as in
\cite[Section~6.1]{berger1971theory}, where the state of a composite source
could vary statistically or be controlled by an adversary.
For a composite source with a statistically varying state, both the encoder and
the decoder could have different knowledge of the switch position, and this led
to different RD functions.
Each of the RD functions was shown to equal the RD function of an appropriately
chosen information source without a state.
If the state is controlled by an adversary, the problem is modeled as the
source coding game.

The composite source model turns out to be very versatile.
Firstly, information sources with uncertain statistical characteristics can be
viewed as mixtures of different sources, i.e. composite sources
\cite[Section~6.1.2]{berger1971theory} \cite{berger1971game, fontana1980}.
Secondly, some complex information sources can be decomposed into simpler
subsources and regarded as composite sources to facilitate analysis.
For example, some sources that are neither discrete nor continuous can be
decomposed into discrete and continuous components so that the behavior of
their RD functions at asymptotically small distortions can be characterized
\cite{rosenthal1988, gyorgy1999}.
Sparse sources can be viewed as composite sources whose subsources tend to
output signals with different magnitudes \cite{weidmann2012}.
A topic beyond our scope here is that non-ergodic sources can be considered as
mixtures of ergodic ones \cite{gray1974}.

The composite source model is also a good model for many real-world sources
such as speech, images, and videos
\cite{kalveram1989, rabiner2007, gibson2014, gibson2017}.
Two classes of composite source models, i.e. hidden Markov and Gaussian mixture
models \cite{rabiner1989}, have been widely adopted in the analysis and
processing of real-world sources.
In our work, each composite source is assumed to have a finite number of
subsources.
As pointed out in \cite{berger1971theory} and in \cite{gray1974}, a generalized
composite source can have uncountably many subsources, and the state of such a
source is a continuous variable.
Generative machine learning models with latent variables, such as generative
adversarial networks \cite{goodfellow2014}, variational autoencoders
\cite{kingma2014}, and diffusion models \cite{croitoru2023}, can be viewed as
generalized composite sources, with latent variables viewed as source states.

Although RD problems on composite sources have been studied extensively,
existing RD analysis does not consider the need to treat subsources of a
composite source differently.
In other words, existing RD analysis does not consider subsource-dependent
fidelity criteria.
Such consideration is important to lossy source coding, and especially to lossy
source coding in semantic communications \cite{gunduz2023}.
In a series of recent works
\cite{kipnis2021, liu2021, liu2022, guo2022, wang2022, xiao2022, stavrou2023},
inference is made on hidden states of information sources, which are regarded
as the meaning or semantics of the sources.
In our work, states can represent the importance of the corresponding output
data.
By putting subsource-dependent fidelity criteria on only the important data, we
can provide another interpretation of semantic communications.

CTC coding is widely adopted in situations where subsource-dependent fidelity
criteria are relevant.
For speech signals, almost every successful coder uses different modes for
different segments of the signals, such as voiced, unvoiced, and silent
segments \cite{wang1989, kalveram1989, gibson2017, das1999}.
Since the segments have different significance to the overall quality of speech
coding, it is beneficial to encode them using different modes, especially for
low-rate speech coding \cite{wang1989, das1999}.
For digital images, the pixels in ROIs comprise a class encoded with higher
quality and the other pixels comprise a class encoded with lower quality.
For example, in JPEG2000, the bits associated with wavelet coefficients in ROIs
can be placed at higher bit-planes than the bits associated with the other
wavelet coefficients \cite{christopoulos2000}.
Consequently, coding of ROIs and the other parts of images can have different
rates, leading to different quality levels.
In remote video analytics systems, video streams can be filtered for
transmission bandwidth savings \cite{wang2023, wang2024}.
The systems can adopt temporal filtering, which selects relevant video frames,
or spatial filtering, which selects relevant pixels from a frame.
Irrelevant frames and pixels are ignored or encoded with low quality so that
the transmission of the filtered video streams does not consume too large
bandwidth.

CTC coding is also adopted for other purposes.
For example, information sources can be modeled as composite sources with
Gaussian subsources.
This model enables designs of robust CTC quantizers with low computational
complexity and variable rates \cite{gray2001, subramaniam2003, zhao2008}.
In spite of its wide applications, information-theoretical evaluation of CTC
coding based on subsource-dependent fidelity criteria is absent up to now.

% cspell: words infimum

\subsection{Notation}
\label{u:introduction.notation}

The set of real numbers is denoted by $\reals$.
For $a$, $b \in \reals \cup \{ - \infty , \infty \}$, $\ocintegers{a}{b}$
denotes the set of integers $n$ satisfying $a < n \le b$, and $\ocnumbers{b}$
denotes a shorthand for $\ocintegers{0}{b}$.
Similarly, $\cointegers{a}{b}$ denotes the set of integers $n$ satisfying
$a \le n < b$, and $\conumbers{b}$ denotes a shorthand for $\cointegers{0}{b}$.
Thus, $\cointegers{1}{\infty}$ denotes the set of positive integers, and
$\conumbers{\infty}$ denotes the set of non-negative integers.
The infimum of the empty set $\emptyset$ is considered as $\infty$.

For a set $J = \{ i_{1} , i_{2} , \cdots , i_{k} \}$ of integers such that
$i_{1} < i_{2} < \cdots < i_{k}$, we let $x_{J}$ denote the sequence
$( x_{i_{1}} , x_{i_{2}} , \cdots , x_{i_{k}} )$.
Thus, $x_{\ocnumbers{n}}$ is the sequence $( x_{1} , x_{2} , \cdots , x_{n} )$.
This sequence is also written as $\{ x_{i} \}_{i = 1}^{n}$ or as the boldface
letter $\boldsymbol{x}$.
The length of a sequence $\boldsymbol{x}$ is denoted by $| \boldsymbol{x} |$.
For two sequences $\boldsymbol{x}$ and $\boldsymbol{y}$,
$\boldsymbol{x} \boldsymbol{y}$ denotes their concatenation.
For any set $A$, the class of all finite sequences composed of the elements of
$A$, including the empty sequence, is denoted by $A^{*}$.

We write $X \markovlink Y \markovlink Z$ if $X$, $Y$, and $Z$ are random
variables and form a Markov chain.
All entropies and mutual information are in units of bits.

    \section{Definitions}
    \label{s:definitions}
    % cspell: words memoryless, subsource, subsources

\subsection{Subsource-Dependent Fidelity Criteria}
\label{u:definitions.criteria}

Let $\mathcal{S}$, $\mathcal{X}$, and $\hat{\mathcal{X}}$ be sets, and
$\mathcal{S}$ be finite.
Suppose random pairs $( S_{1} , X_{1} )$, $( S_{2} , X_{2} )$, $\cdots$ are
independent and identically distributed (i.i.d.) and take values in
$\mathcal{S} \times \mathcal{X}$.
We refer to $\{ ( S_{i} , X_{i} ) \}_{i = 1}^{\infty}$ as a memoryless
composite source.
For $i \in \cointegers{1}{\infty}$, $X_{i}$ is an output symbol and is thought
to be produced by Subsource~$S_{i}$.
Let $\{ s_{i} \}_{i = 1}^{n} \in \mathcal{S}^{n}$ and
$\{ x_{i} \}_{i = 1}^{n} \in \mathcal{X}^{n}$ be realizations of
$\{ S_{i} \}_{i = 1}^{n}$ and $\{ X_{i} \}_{i = 1}^{n}$, respectively.
Suppose $\{ x_{i} \}_{i = 1}^{n}$ is reproduced as
$\{ \hat{x}_{i} \}_{i = 1}^{n} \in \hat{\mathcal{X}}^{n}$.

A classical fidelity criterion requires
\begin{equation}
    \frac{1}{n} \sum_{i = 1}^{n} d ( x_{i} , \hat{x}_{i} ) \le \delta ,
    \label{e:definitions.criteria.classical}
\end{equation}
where $\delta \in \cointerval{0}{\infty}$ and $d$ is a distortion measure on
$\mathcal{X} \times \hat{\mathcal{X}}$, i.e. a function from
$\mathcal{X} \times \hat{\mathcal{X}}$ to $\cointerval{0}{\infty}$.
For $x \in \mathcal{X}$ and $\hat{x} \in \hat{\mathcal{X}}$,
$d ( x , \hat{x} )$ is the distortion level of $\hat{x}$ as a reproduction of
$x$.
The larger $d ( x , \hat{x} )$, the lower the fidelity.
Unlike \eqref{e:definitions.criteria.classical}, the subsource-dependent
fidelity criterion involves the state sequence $s_{\ocnumbers{n}}$ and a subset
$\mathcal{S}_{*}$ of $\mathcal{S}$, in addition to $x_{\ocnumbers{n}}$,
$\hat{x}_{\ocnumbers{n}}$, $d$, and $\delta$.
We say $s_{\ocnumbers{n}}$, $x_{\ocnumbers{n}}$, and $\hat{x}_{\ocnumbers{n}}$
meet the fidelity criterion $( d , \mathcal{S}_{*} , \delta )$ if
\begin{equation}
    \sum_{i \in \ocnumbers{n} , s_{i} \in \mathcal{S}_{*}}
    ( d ( x_{i} , \hat{x}_{i} ) - \delta )
    \le 0 .
    \label{e:definitions.criteria.ssdependent} % subsource-dependent
\end{equation}
When the number $m = | \{ i \in \ocnumbers{n} | s_{i} \in \mathcal{S}_{*} \} |$
of terms satisfying $s_{i} \in \mathcal{S}_{*}$ is non-zero,
\eqref{e:definitions.criteria.ssdependent} can be written as
\begin{equation}
    \frac{1}{m}
    \sum_{i \in \ocnumbers{n} , s_{i} \in \mathcal{S}_{*}}
    d ( x_{i} , \hat{x}_{i} )
    \le \delta ,
\end{equation}
implying that the average distortion of the symbols from the subsources in
$\mathcal{S}_{*}$ does not exceed $\delta$.
If $\mathcal{S}_{*} = \mathcal{S}$, the fidelity criterion
$( d , \mathcal{S}_{*} , \delta )$ is equivalent to the classical fidelity
criterion.
If $\mathcal{S}_{*} \not= \mathcal{S}$, the criterion
$( d , \mathcal{S}_{*} , \delta )$ measures fidelity of the symbols from the
subsources in $\mathcal{S}_{*}$.

We will consider a family $
    \{
        ( d_{\lambda} , \mathcal{S}_{\lambda} , D ( \lambda ) )
    \}_{\lambda \in \Lambda}$
of fidelity criteria indexed by a finite set $\Lambda$, where for
$\lambda \in \Lambda$, $d_{\lambda}$ is a distortion measure on
$\mathcal{X} \times \hat{\mathcal{X}}$, $\mathcal{S}_{\lambda}$ is a subset of
$\mathcal{S}$, and $D ( \lambda ) \in \cointerval{0}{\infty}$ is a distortion
level.
For example, we can let $\Lambda = \mathcal{S}$ and
$\mathcal{S}_{\lambda} = \{ \lambda \}$ for $\lambda \in \Lambda$.
Then the fidelity criteria $\{
    ( d_{\lambda} , \mathcal{S}_{\lambda} , D ( \lambda ) )
\}_{\lambda \in \Lambda}$ require that the average distortion of symbols from
Subsource~$s$ does not exceed $D ( s )$, for all $s \in \mathcal{S}$.
We can define $\{ \mathcal{S}_{\lambda} \}_{\lambda \in \Lambda}$ in a more
flexible fashion to describe many other fidelity requirements.

    % cspell: words losslessly

\subsection{Variable-Length Codes and CTC Coding}
\label{u:definitions.codes}

In RD theory, a frequently used model for lossy source coding is a block code.
A block code can be defined \cite{cover2006, csiszar2011} as a pair
$( f , \varphi )$ of functions $f : \mathcal{X}^{n} \to \ocnumbers{M}$ and
$\varphi : \ocnumbers{M} \to \hat{\mathcal{X}}^{n}$, where $M$ and $n$ are
positive integers.
The function $f$ describes an encoder that maps a sequence
$\boldsymbol{x} \in \mathcal{X}^{n}$ of source symbols to an index
$m = f ( \boldsymbol{x} )$.
The function $\varphi$ describes a decoder that accepts $m$ as an input and
reproduces $\boldsymbol{x}$ as $\varphi ( m ) \in \hat{\mathcal{X}}^{n}$.
A block code is a convenient model if the number of source symbols to be
compressed is fixed.
This is not the case in CTC coding, where the number of symbols in a class is
random.
Hence, we use variable-length codes in our analysis of CTC coding.
We also use variable-length codes to describe other results for uniformity.

\begin{definition}
    \label{def:definitions.codes.varlen} % variable-length
    An $( \mathcal{X} , \hat{\mathcal{X}} )$-variable-length code is a pair
    $( f , \varphi )$ of functions $f : \mathcal{X}^{*} \to \{ 0 , 1 \}^{*}$
    and $\varphi : \{ 0 , 1 \}^{*} \to \hat{\mathcal{X}}^{*}$ satisfying
    $| \varphi ( f ( \boldsymbol{x} ) ) | = | \boldsymbol{x} |$ for all
    $\boldsymbol{x} \in \mathcal{X}^{*}$.
\end{definition}

\begin{definition}
    \label{def:definitions.codes.lossless}
    An $( \mathcal{X} , \mathcal{X} )$-variable-length code is referred to as
    an $\mathcal{X}$-lossless code if
    $\varphi ( f ( \boldsymbol{x} ) ) = \boldsymbol{x}$ for all
    $\boldsymbol{x} \in \mathcal{X}^{*}$.
\end{definition}

In the $( \mathcal{X} , \hat{\mathcal{X}} )$-variable-length code
$( f , \varphi )$, $f$ describes how an encoder maps the sequence
$\boldsymbol{x}$ of source symbols to the bit string $f ( \boldsymbol{x} )$,
and $\varphi$ describes how a decoder maps the bit string
$f ( \boldsymbol{x} )$ to the sequence $\varphi ( f ( \boldsymbol{x} ) )$ of
reproduced symbols.
A key requirement is that the reproduced sequence
$\varphi ( f ( \boldsymbol{x} ) )$ has the same length as the original sequence
$\boldsymbol{x}$ of source symbols.

\begin{figure*}[!t]
    \centering \footnotesize
    % cspell: words bitstring

\begin{tikzpicture}[
    bit/.style = {
        draw,
        inner sep = 0,
        minimum height = 1ex,
        minimum width = 1ex
    },
    longblock/.style = {draw, minimum height = 4ex, minimum width = 6em},
    shortblock/.style = {draw, minimum height = 4ex, minimum width = 2em},
    symbol/.style = {
        draw,
        inner sep = 0,
        minimum height = 2ex,
        minimum width = 1ex
    }
]
    % 分类器, 编码器, 译码器和一些坐标
    \matrix[row sep = 2ex, column sep = 2.3em] {
        \coordinate (source); &
        \node[longblock] (classifier) {Classifier $c$}; &
        \coordinate (labels); &
        \node[shortblock] (label encoder) {$\tilde{f}$}; &
        &
        \node[shortblock] (label decoder) {$\tilde{\varphi}$}; &
        \coordinate (reproduced labels); &
        \node[longblock] (multiplexer) {Multiplexer}; &
        \coordinate (reproduced); \\
        &
        &
        \coordinate (symbols 1); &
        \node[shortblock] (encoder 1) {$f_{1}$}; &
        \coordinate (bit string); &
        \node[shortblock] (decoder 1) {$\varphi_{1}$}; &
        \coordinate (reproduced 1); &
        &
        \\
        &
        &
        \coordinate (symbols 2); &
        \node[shortblock] (encoder 2) {$f_{2}$}; &
        &
        \node[shortblock] (decoder 2) {$\varphi_{2}$}; &
        \coordinate (reproduced 2); &
        &
        \\
    };

    % 信源符号序列
    \begin{scope}[shift = (source), x = 1ex, y = 2ex]
        \foreach \x in {-2, 1} {
            \node[symbol, fill = cyan] at (\x, 0) {};
        }
        \foreach \x in {-1, 0, 2} {
            \node[symbol, fill = orange] at (\x, 0) {};
        }
        \node[above] at (0, 0.5) {$x_{\ocnumbers{n}}$};
        \draw[->] (2.5, 0) -- (classifier);
    \end{scope}

    % 类别组成的序列
    \begin{scope}[shift = (labels), x = 1ex, y = 2ex]
        \foreach \x in {-2, -1, 0, 1, 2} {
            \node[symbol] at (\x, 0) {};
        }
        \node[above] at (0, 0.5) {$u_{\ocnumbers{n}}$};
        \draw[->] (classifier) -- (-2.5, 0);
        \draw[->] (2.5, 0) -- (label encoder);
    \end{scope}

    % 第 1 类信源符号
    \begin{scope}[shift = (symbols 1), x = 1ex, y = 2ex]
        \foreach \x in {-0.5, 0.5} {
            \node[symbol, fill = cyan] at (\x, 0) {};
        }
        \node[above] at (0, 0.5) {$x_{J ( 1 )}$};
        \draw[->] ($ (classifier.south) + (1em, 0) $) |- (-1, 0);
        \draw[->] (1, 0) -- (encoder 1);
    \end{scope}

    % 第 2 类信源符号
    \begin{scope}[shift = (symbols 2), x = 1ex, y = 2ex]
        \foreach \x in {-1, 0, 1} {
            \node[symbol, fill = orange] at (\x, 0) {};
        }
        \node[above] at (0, 0.5) {$x_{J ( 2 )}$};
        \draw[->] ($ (classifier.south) + (-1em, 0) $) |- (-1.5, 0);
        \draw[->] (1.5, 0) -- (encoder 2);
    \end{scope}

    % 比特串
    \begin{scope}[shift = (bit string), x = 1ex, y = 1ex]
        \foreach \y in {7.5, 6.5, 5.5, 4.5} {
            \node[bit] at (0, \y) {};
        }
        \draw[->] (label encoder) -- (label encoder -| -0.5, 0)
        node[above left] {$\tilde{\boldsymbol{b}}$};
        \draw[->] (0.5, 0 |- label decoder) -- (label decoder);
        \foreach \y in {3.5, 2.5, ..., -0.5} {
            \node[bit, fill = cyan] at (0, \y) {};
        }
        \draw[->] (encoder 1) -- (-0.5, 0) node[above left]
        {$\boldsymbol{b}_{1}$};
        \draw[->] (0.5, 0) -- (decoder 1);
        \foreach \y in {-1.5, -2.5, ..., -7.5} {
            \node[bit, fill = orange] at (0, \y) {};
        }
        \draw[->] (encoder 2) -- (encoder 2 -| -0.5, 0) node[above left]
        {$\boldsymbol{b}_{2}$};
        \draw[->] (0.5, 0 |- decoder 2) -- (decoder 2);
    \end{scope}

    % 恢复的类别序列
    \begin{scope}[shift = (reproduced labels), x = 1ex, y = 2ex]
        \foreach \x in {-2, -1, ..., 2} {
            \node[symbol] at (\x, 0) {};
        }
        \node[above] at (0, 0.5) {$u_{\ocnumbers{n}}$};
        \draw[->] (label decoder) -- (-2.5, 0);
        \draw[->] (2.5, 0) -- (multiplexer);
    \end{scope}

    % 恢复的第 1 类信源符号
    \begin{scope}[shift = (reproduced 1), x = 1ex, y = 2ex]
        \foreach \x in {-0.5, 0.5} {
            \node[symbol, fill = cyan] at (\x, 0) {};
        }
        \node[above] at (0, 0.5) {$\hat{x}_{J ( 1 )}$};
        \draw[->] (decoder 1) -- (-1, 0);
        \draw[->] (1, 0) -| ($ (multiplexer.south) + (-1em, 0) $);
    \end{scope}

    % 恢复的第 2 类信源符号
    \begin{scope}[shift = (reproduced 2), x = 0.5em, y = 2ex]
        \foreach \x in {-1, 0, 1} {
            \node[symbol, fill = orange] at (\x, 0) {};
        }
        \node[above] at (0, 0.5) {$\hat{x}_{J ( 2 )}$};
        \draw[->] (decoder 2) -- (-1.5, 0);
        \draw[->] (1.5, 0) -| ($ (multiplexer.south) + (1em, 0) $);
    \end{scope}

    % 恢复的信源符号序列
    \begin{scope}[shift = (reproduced), x = 0.5em, y = 2ex]
        \foreach \x in {-2, 1} {
            \node[symbol, fill = cyan] at (\x, 0) {};
        }
        \foreach \x in {-1, 0, 2} {
            \node[symbol, fill = orange] at (\x, 0) {};
        }
        \node[above] at (0, 0.5) {$\hat{x}_{\ocnumbers{n}}$};
        \draw[->] (multiplexer) -- (-2.5, 0);
    \end{scope}
\end{tikzpicture}
    \begin{tikzpicture}[
    bit/.style = {
        draw,
        inner sep = 0,
        minimum height = 1ex,
        minimum width = 1ex
    },
    symbol/.style = {
        draw,
        inner sep = 0,
        minimum height = 2ex,
        minimum width = 1ex
    }
]
    \matrix[column sep = 1em] {
        \node[symbol] at (0, 0) {}; &
        \node[right] at (0, 0) {Class label}; \\
        \node[symbol, fill = cyan] at (0, 0) {}; &
        \node[right] at (0, 0) {Symbol in Class~$1$}; \\
        \node[symbol, fill = orange] at (0, 0) {}; &
        \node[right] at (0, 0) {Symbol in Class~$2$}; \\
        \node[bit] at (-2em, 0) {};
        \node[bit, fill = cyan] at (-1em, 0) {};
        \node[bit, fill = orange] at (0, 0) {}; &
        \node[right] at (0, 0) {Bits}; \\
    };
\end{tikzpicture}
    \caption{An example of CTC coding.}
    \label{f:definitions.codes.labelctc}
\end{figure*}
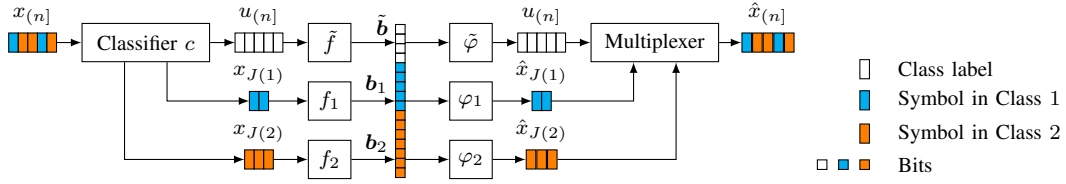

CTC codes are variable-length codes that accomplish CTC coding.
Before defining CTC codes formally, let us consider an example of CTC coding.
Let $L$ be a positive integer, $\mathcal{U} = \ocnumbers{L}$, and
$c : \mathcal{X} \to \mathcal{U}$ be a classifier that assigns the label
$c ( x )$ to each $x \in \mathcal{X}$.
Define $u_{i} = c ( x_{i} )$ for $i \in \ocnumbers{n}$ and
$J ( u ) = \{ i \in \ocnumbers{n} | u_{i} = u \}$ for $u \in \mathcal{U}$.
An encoder encodes $u_{\ocnumbers{n}}$ as a bit string
$\tilde{\boldsymbol{b}} = \tilde{f} ( u_{\ocnumbers{n}} )$ using a
$\mathcal{U}$-lossless code $( \tilde{f} , \tilde{\varphi} )$.
The encoder also encodes the sequence $x_{J ( u )}$ of the symbols in Class~$u$
as a bit string $\boldsymbol{b}_{u} = f_{u} ( x_{J ( u )} )$ using an
$( \mathcal{X} , \hat{\mathcal{X}} )$-variable-length code
$( f_{u} , \varphi_{u} )$.
The encoder then outputs the concatenated bit string $
    \tilde{\boldsymbol{b}}
    \boldsymbol{b}_{1}
    \boldsymbol{b}_{2}
    \cdots
    \boldsymbol{b}_{L}$.
A decoder recovers $\tilde{\boldsymbol{b}}$, $\boldsymbol{b}_{1}$,
$\boldsymbol{b}_{2}$, $\cdots$, $\boldsymbol{b}_{L}$ from the concatenated bit
string, reproduces $u_{\ocnumbers{n}}$ as
$\tilde{\varphi} ( \tilde{\boldsymbol{b}} )$, and reproduces $x_{J ( u )}$ as
$\varphi_{u} ( \boldsymbol{b}_{u} )$ for all $u \in \mathcal{U}$.
Because $( \tilde{f} , \tilde{\varphi} )$ is lossless,
$\tilde{\varphi} ( \tilde{\boldsymbol{b}} )$ equals $u_{\ocnumbers{n}}$.
The decoder replaces the $u$'s in $u_{\ocnumbers{n}}$ with
$\varphi_{u} ( \boldsymbol{b}_{u} )$ for all $u \in \mathcal{U}$ to get
$\{ \hat{x}_{i} \}_{i = 1}^{n}$ as the reproduced $x_{\ocnumbers{n}}$.
The process for $L = 2$ is shown in
\figurename~\ref{f:definitions.codes.labelctc}.
We can see $\hat{x}_{J ( u )} = \varphi_{u} ( f_{u} ( x_{J ( u )} ) )$ for
$u \in \mathcal{U}$.

\begin{figure}[t]
    \centering \footnotesize
    \begin{tikzpicture}[
    bit/.style = {
        draw,
        inner sep = 0,
        minimum height = 1ex,
        minimum width = 1ex
    },
    x = 1ex,
    y = 1ex
]
    \foreach \x in {0, 1, 2} {
        \node[bit] at (\x, 0) {};
    }
    \node[above] at (1, 0.5) {$\tilde{\boldsymbol{b}}_{1}$};
    \foreach \x in {3, 4, ..., 7} {
        \node[bit, fill = cyan] at (\x, 0) {};
    }
    \node[above] at (5, 0.5) {$\boldsymbol{b}_{1}$};
    \foreach \x in {8, 9, 10, 11} {
        \node[bit] at (\x, 0) {};
    }
    \node[above] at (9.5, 0.5) {$\tilde{\boldsymbol{b}}_{2}$};
    \foreach \x in {12, 13, ..., 18} {
        \node[bit, fill = orange] at (\x, 0) {};
    }
    \node[above] at (15, 0.5) {$\boldsymbol{b}_{2}$};
    \foreach \x in {19, 20, 21} {
        \node[bit] at (\x, 0) {};
    }
    \node[above] at (20, 0.5) {$\tilde{\boldsymbol{b}}_{3}$};
    \foreach \x in {22, 23, ..., 27} {
        \node[bit, fill = green] at (\x, 0) {};
    }
    \node[above] at (24.5, 0.5) {$\boldsymbol{b}_{3}$};
\end{tikzpicture}
    \caption{%
        The transmitted bit string in another implementation of CTC coding with
        $\mathcal{U} = \ocnumbers{3}$.%
    }
    \label{f:definitions.codes.positionctc}
\end{figure}
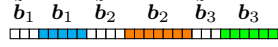

CTC coding can be implemented in ways different from that shown in
\figurename~\ref{f:definitions.codes.labelctc}.
For example, we can insert a bit string $\tilde{\boldsymbol{b}}_{u}$ before
$\boldsymbol{b}_{u}$ to indicate which symbols in $x_{\ocnumbers{n}}$ are in
Class~$u$, as shown in \figurename~\ref{f:definitions.codes.positionctc}.
Then the bit string $\tilde{\boldsymbol{b}}$ representing labels can be
omitted.
It is not difficult to think of other implementations of CTC coding.
CTC coding has the following characteristics no matter how it is implemented:
\begin{enumerate}
    \item The sequence $u_{\ocnumbers{n}}$ of labels can be correctly recovered
    from the transmitted bit string; and
    \item The classes are compressed separately, that is,
    $\hat{x}_{J ( u )} = \varphi_{u} ( f_{u} ( x_{J ( u )} ) )$ for all
    $x_{\ocnumbers{n}}$ and $u$.
\end{enumerate}
We use these characteristics to define CTC codes.
In general, the label set $\mathcal{U}$ need not be $\ocnumbers{L}$, and can be
an arbitrary finite set.

\begin{definition}
    \label{def:definitions.codes.ctc}
    An $( \mathcal{X} , \hat{\mathcal{X}} )$-variable-length code
    $( f , \varphi )$ is referred to as an
    $( \mathcal{X} , \hat{\mathcal{X}} , \mathcal{U} , c )$-CTC code if there
    exist functions $\psi : \{ 0 , 1 \}^{*} \to \mathcal{U}^{*}$ and
    $g : \mathcal{X}^{*} \times \mathcal{U} \to \hat{\mathcal{X}}^{*}$ such
    that
    \begin{enumerate}
        \item
        $\psi ( f ( x_{\ocnumbers{n}} ) ) = \{ c ( x_{i} ) \}_{i = 1}^{n}$ for
        all $n \in \conumbers{\infty}$ and
        $\{ x_{i} \}_{i = 1}^{n} \in \mathcal{X}^{n}$; and
        \item If $n \in \conumbers{\infty}$ and $
            \{ \hat{x}_{i} \}_{i = 1}^{n}
            = \varphi ( f ( \{ x_{i} \}_{i = 1}^{n} ) )
        $, then $\hat{x}_{J ( u )} = g ( x_{J ( u )} , u )$ for all
        $u \in \mathcal{U}$, where
        $J ( u ) = \{ i \in \ocnumbers{n} | c ( x_{i} ) = u \}$.
    \end{enumerate}
\end{definition}

For $u \in \mathcal{U}$, $g ( \cdot , u )$ describes how the symbols in
Class~$u$ are encoded and decoded, like $\varphi_{u} ( f_{u} ( \cdot ) )$ in
\figurename~\ref{f:definitions.codes.labelctc}.

The CTC code illustrated in \figurename~\ref{f:definitions.codes.labelctc} can
be viewed as the combination of two codes: One compresses the labels
losslessly, and the other compresses the source symbols based on their labels.
We refer to the latter code as a label-based code.

\begin{definition}
    \label{def:definitions.codes.label}
    An $( \mathcal{X} , \hat{\mathcal{X}} , \mathcal{U} )$-label-based code is
    a pair $( f , \varphi )$ of functions
    $f : ( \mathcal{X} \times \mathcal{U} )^{*} \to \{ 0 , 1 \}^{*}$ and $
        \varphi :
        \{ 0 , 1 \}^{*} \times \mathcal{U}^{*} \to \hat{\mathcal{X}}^{*}$
    satisfying $
        | \varphi ( f ( \boldsymbol{x} , \boldsymbol{u} ) , \boldsymbol{u} ) |
        = n$
    for all $n \in \conumbers{\infty}$, $\boldsymbol{x} \in \mathcal{X}^{n}$,
    and $\boldsymbol{u} \in \mathcal{U}^{n}$.
\end{definition}

In Definition~\ref{def:definitions.codes.label}, we require the original
sequence $\boldsymbol{x}$, the reproduced sequence
$\varphi ( f ( \boldsymbol{x} , \boldsymbol{u} ) , \boldsymbol{u} )$, and the
label sequence $\boldsymbol{u}$ to have the same length, because every source
symbol has a label.

    \subsection{RD Functions}
\label{u:definitions.rdfunctions} % rate-distortion functions

Let $( S , X )$ be a random pair such that $( S , X )$, $( S_{1} , X_{1} )$,
$( S_{2} , X_{2} )$, $\cdots$ are identically distributed.
Define $
    \bar{\Lambda}
    = \{ \lambda \in \Lambda | \probability{S \in \mathcal{S}_{\lambda}} > 0 \}
$.
For an arbitrary family $\{ D ( \lambda ) \}_{\lambda \in \Lambda}$ of
distortion levels, let
\begin{gather}
    R^{*} ( D )
    = \inf_{\hat{X} \in \Xi ( D )} \minfo{X ; \hat{X}} ,
    \label{e:results.rate} \\
    R^{\mathrm{C}} ( D )
    = \inf_{\hat{X} \in \Xi ( D )} \minfo{X ; \hat{X} , c ( X )} ,
    \label{e:results.ctcrate}
\end{gather}
and
\begin{equation}
    R^{\mathrm{G}} ( D )
    = \inf_{\hat{X} \in \Xi ( D )} \minfo{X ; \hat{X} | c ( X )} ,
    \label{e:results.parate} % parallel rate
\end{equation}
where $\Xi ( D )$ is the set of random variables $\hat{X}$ satisfying
$S \markovlink X \markovlink \hat{X}$ and $
    \mean{d_{\lambda} ( X , \hat{X} ) | S \in \mathcal{S}_{\lambda}}
    \le D ( \lambda )
$ for all $\lambda \in \bar{\Lambda}$.
Equation~\eqref{e:results.rate} defines the RD function $R^{*}$ of $( S , X )$
for $\{ ( d_{\lambda} , \mathcal{S}_{\lambda} ) \}_{\lambda \in \Lambda}$.
Equation~\eqref{e:results.ctcrate} defines the RD function $R^{\mathrm{C}}$ of
$( S , X )$ for
$\{ ( d_{\lambda} , \mathcal{S}_{\lambda} ) \}_{\lambda \in \Lambda}$ and
$( \mathcal{X} , \hat{\mathcal{X}} , \mathcal{U} , c )$-CTC codes.
Equation~\eqref{e:results.parate} defines the RD function $R^{\mathrm{G}}$ of
$( S , X )$ for
$\{ ( d_{\lambda} , \mathcal{S}_{\lambda} ) \}_{\lambda \in \Lambda}$ given
$c ( X )$.
In the next section, we will see that $R^{*}$, $R^{\mathrm{C}}$, and
$R^{\mathrm{G}}$ characterize the optimal performance of
$( \mathcal{X} , \hat{\mathcal{X}} )$-variable-length codes,
$( \mathcal{X} , \hat{\mathcal{X}} , \mathcal{U} , c )$-CTC codes, and
$( \mathcal{X} , \hat{\mathcal{X}} , \mathcal{U} )$-label-based codes,
respectively.

    \section{Main Results}
    \label{s:results}
    % cspell: words achievability

\subsection{Coding Theorems}
\label{u:results.theorems}

Fix $R \in \cointerval{0}{\infty}$ and a family
$\{ D ( \lambda ) \}_{\lambda \in \Lambda}$ of distortion levels.
An $( \mathcal{X} , \hat{\mathcal{X}} )$-variable-length code is said to
achieve $( R , D )$ for $( S , X )$ and
$\{ ( d_{\lambda} , \mathcal{S}_{\lambda} ) \}_{\lambda \in \Lambda}$ if
\begin{equation}
    \lim_{n \to \infty}
    \probability{( S_{\ocnumbers{n}} , X_{\ocnumbers{n}} ) \in F_{\epsilon}}
    = 1
    \label{e:results.theorems.vla} % variable-length achievability
\end{equation}
for all $\epsilon \in \oointerval{0}{\infty}$, where $F_{\epsilon}$ is the set
of $
    ( \boldsymbol{s} , \boldsymbol{x} )
    \in ( \mathcal{S} \times \mathcal{X} )^{*}
$ such that $| f ( \boldsymbol{x} ) | \le | \boldsymbol{x} | R$ and such that
$\boldsymbol{s}$, $\boldsymbol{x}$, and $\varphi ( f ( \boldsymbol{x} ) )$ meet
the fidelity criterion
$( d_{\lambda} , \mathcal{S}_{\lambda} , D ( \lambda ) + \epsilon )$ for all
$\lambda \in \Lambda$.
Define $U_{i} = c ( X_{i} )$ for $i \in \cointegers{1}{\infty}$.
An $( \mathcal{X} , \hat{\mathcal{X}} , \mathcal{U} )$-label-based code is said
to achieve $( R , D )$ for $( S , X )$ and
$\{ ( d_{\lambda} , \mathcal{S}_{\lambda} ) \}_{\lambda \in \Lambda}$ given
$c ( X )$ if
\begin{equation}
    \lim_{n \to \infty} \probability{
        ( S_{\ocnumbers{n}} , X_{\ocnumbers{n}} , U_{\ocnumbers{n}} )
        \in G_{\epsilon}
    }
    = 1
\end{equation}
for all $\epsilon \in \oointerval{0}{\infty}$, where $G_{\epsilon}$ is the set
of $
    ( \boldsymbol{s} , \boldsymbol{x} , \boldsymbol{u} )
    \in ( \mathcal{S} \times \mathcal{X} \times \mathcal{U} )^{*}
$ such that
$| f ( \boldsymbol{x} , \boldsymbol{u} ) | \le | \boldsymbol{x} | R$ and such
that $\boldsymbol{s}$, $\boldsymbol{x}$, and
$\varphi ( f ( \boldsymbol{x} , \boldsymbol{u} ) , \boldsymbol{u} )$ meet the
fidelity criterion
$( d_{\lambda} , \mathcal{S}_{\lambda} , D ( \lambda ) + \epsilon )$ for all
$\lambda \in \Lambda$.

The following coding theorems are valid when $\mathcal{X}$ and
$\hat{\mathcal{X}}$ are finite.
They may be generalized to abstract sources, but with some additional technical
conditions \cite{berger1971theory, gray2023}.

\begin{theorem}
    \label{thm:results.theorems.general}
    If $R > R^{*} ( D )$, then there exists an
    $( \mathcal{X} , \hat{\mathcal{X}} )$-variable-length code
    $( f , \varphi )$ achieving $( R , D )$ for $( S , X )$ and
    $\{ ( d_{\lambda} , \mathcal{S}_{\lambda} ) \}_{\lambda \in \Lambda}$, and
    satisfying
    \begin{equation}
        \limsup_{n \to \infty}
        \frac{1}{n}
        \max_{\boldsymbol{x} \in \mathcal{X}^{n}}
        | f ( \boldsymbol{x} ) |
        < R . \label{e:results.theorems.rate}
    \end{equation}
    Conversely, if an $( \mathcal{X} , \hat{\mathcal{X}} )$-variable-length
    code $( f , \varphi )$ achieves $( R , D )$ for $( S , X )$ and
    $\{ ( d_{\lambda} , \mathcal{S}_{\lambda} ) \}_{\lambda \in \Lambda}$, then
    $R \ge R^{*} ( D )$.
\end{theorem}

\begin{theorem}
    \label{thm:results.theorems.ctc}
    If $R > R^{\mathrm{C}} ( D )$, then there exists an
    $( \mathcal{X} , \hat{\mathcal{X}} , \mathcal{U} , c )$-CTC code
    achieving\footnote{%
        This means that \eqref{e:results.theorems.vla} holds for all
        $\epsilon \in \oointerval{0}{\infty}$, because any CTC code is a
        variable-length code.%
    } $( R , D )$ for $( S , X )$ and
    $\{ ( d_{\lambda} , \mathcal{S}_{\lambda} ) \}_{\lambda \in \Lambda}$.
    Conversely, if an $( \mathcal{X} , \hat{\mathcal{X}} )$-variable-length
    code $( f , \varphi )$ achieves $( R , D )$ for $( S , X )$ and
    $\{ ( d_{\lambda} , \mathcal{S}_{\lambda} ) \}_{\lambda \in \Lambda}$,
    $\psi$ is a function from $\{ 0 , 1 \}^{*}$ to $\mathcal{U}^{*}$, and
    $\psi ( f ( X_{\ocnumbers{n}} ) ) = U_{\ocnumbers{n}}$ with probability one
    for all $n \in \conumbers{\infty}$, then $R \ge R^{\mathrm{C}} ( D )$.
\end{theorem}

\begin{theorem}
    \label{thm:results.theorems.condition}
    If $R > R^{\mathrm{G}} ( D )$, then there exist an
    $( \mathcal{X} , \hat{\mathcal{X}} , \mathcal{U} )$-label-based code
    $( f , \varphi )$ and a function
    $g : \mathcal{X}^{*} \times \mathcal{U} \to \hat{\mathcal{X}}^{*}$ such
    that
    \begin{enumerate}
        \item $( f , \varphi )$ achieves $( R , D )$ for $( S , X )$ and
        $\{ ( d_{\lambda} , \mathcal{S}_{\lambda} ) \}_{\lambda \in \Lambda}$
        given $c ( X )$; and
        \item If $n \in \conumbers{\infty}$ and
        \begin{equation}
            \{ \hat{x}_{i} \}_{i = 1}^{n}
            = \varphi (
                f ( \{ x_{i} \}_{i = 1}^{n} , \{ u_{i} \}_{i = 1}^{n} ) ,
                \{ u_{i} \}_{i = 1}^{n}
            ) , \label{e:results.theorems.sicoding} % side information coding
        \end{equation}
        then $\hat{x}_{J ( u )} = g ( x_{J ( u )} , u )$ for all
        $u \in \mathcal{U}$, where
        $J ( u ) = \{ i \in \ocnumbers{n} | u_{i} = u \}$.
    \end{enumerate}
    Conversely, if there exists an
    $( \mathcal{X} , \hat{\mathcal{X}} , \mathcal{U} )$-label-based code
    $( f , \varphi )$ achieving $( R , D )$ for $( S , X )$ and
    $\{ ( d_{\lambda} , \mathcal{S}_{\lambda} ) \}_{\lambda \in \Lambda}$ given
    $c ( X )$, then $R \ge R^{\mathrm{G}} ( D )$.
\end{theorem}

The theorems say that $( \mathcal{X} , \hat{\mathcal{X}} )$-variable-length
codes, $( \mathcal{X} , \hat{\mathcal{X}} , \mathcal{U} , c )$-CTC codes, and
$( \mathcal{X} , \hat{\mathcal{X}} , \mathcal{U} )$-label-based codes need
rates no less than $R^{*} ( D )$, $R^{\mathrm{C}} ( D )$, and
$R^{\mathrm{G}} ( D )$, respectively, for the fidelity criteria $\{
    ( d_{\lambda} , \mathcal{S}_{\lambda} , D ( \lambda ) )
\}_{\lambda \in \Lambda}$.
The converse part of Theorem~\ref{thm:results.theorems.ctc} does not need the
condition that $( f , \varphi )$ is an
$( \mathcal{X} , \hat{\mathcal{X}} , \mathcal{U} , c )$-CTC code.
It only needs the condition that class labels $U_{\ocnumbers{n}}$ can be
correctly recovered from the transmitted bit string $f ( X_{\ocnumbers{n}} )$,
i.e. the first characteristic of CTC codes (see the paragraph preceding
Definition~\ref{def:definitions.codes.ctc}).
Hence, an $( \mathcal{X} , \hat{\mathcal{X}} , \mathcal{U} , c )$-CTC code
cannot perform better even if it is allowed to compress symbols in different
classes jointly.
In other words, the performance loss of CTC coding relative to RD-optimal
coding is not due to separate compression of source symbols in different
classes.
The performance loss is because class labels have to be reproduced in CTC
coding.
The existence of the function $g$ in the direct part of
Theorem~\ref{thm:results.theorems.condition} means that $( f , \varphi )$
compresses symbols in different classes separately.
Hence, separate compression does not cause performance loss for label-based
codes either.

    % cspell: words Arimoto, Blahut, minimizations, subsource, subsources

\subsection{Properties of RD Functions}
\label{u:results.properties}

For an arbitrary random variable $\hat{X}$, we have
\begin{gather}
    \minfo{X ; \hat{X} | c ( X )} + \minfo{c ( X ) ; \hat{X}}
    = \minfo{X ; \hat{X}} ,
    \label{e:results.properties.igai} \\ % information gap as information
    \minfo{X ; \hat{X}} + \entropy{c ( X ) | \hat{X}}
    = \minfo{X ; \hat{X} , c ( X )} ,
    \label{e:results.properties.igae} % information gap as entropy
\end{gather}
and
\begin{equation}
    \minfo{X ; \hat{X} | c ( X )} + \entropy{c ( X )}
    = \minfo{X ; \hat{X} , c ( X )} ,
\end{equation}
leading to the following corollary.

\begin{corollary}
    \label{cor:results.properties.order}
    For every family $\{ D ( \lambda ) \}_{\lambda \in \Lambda}$ of distortion
    levels, $
        R^{\mathrm{G}} ( D ) \le R^{*} ( D ) \le R^{\mathrm{C}} ( D )
        = R^{\mathrm{G}} ( D ) + \entropy{c ( X )}
    $.
\end{corollary}

The identity $R^{\mathrm{C}} ( D ) = R^{\mathrm{G}} ( D ) + \entropy{c ( X )}$
has a straightforward interpretation that CTC codes need to compress labels in
addition to source symbols.

\begin{corollary}
    \label{cor:results.properties.classical}
    Define distortion measures
    $\{ \hat{d}_{\lambda} \}_{\lambda \in \bar{\Lambda}}$ on
    $\mathcal{X} \times \hat{\mathcal{X}}$ by
    \begin{equation}
        \hat{d}_{\lambda} ( x , \hat{x} )
        = \frac{
            \probability{S \in \mathcal{S}_{\lambda} | X = x}
        }{
            \probability{S \in \mathcal{S}_{\lambda}}
        }
        d_{\lambda} ( x , \hat{x} ) .
    \end{equation}
    Then for every $\{ D ( \lambda ) \}_{\lambda \in \Lambda}$,
    \begin{equation}
        R^{*} ( D )
        = \inf_{\hat{X} \in \hat{\Xi} ( D )} \minfo{X ; \hat{X}} ,
        \label{e:results.properties.classical}
    \end{equation}
    where $\hat{\Xi} ( D )$ is the set of random variables $\hat{X}$ satisfying
    $\mean{\hat{d}_{\lambda} ( X , \hat{X} )} \le D ( \lambda )$ for all
    $\lambda \in \bar{\Lambda}$.
\end{corollary}

We can prove Corollary~\ref{cor:results.properties.classical} by noting that
\begin{equation}
    \mean{\hat{d}_{\lambda} ( X , \hat{X} )}
    = \mean{d_{\lambda} ( X , \hat{X} ) | S \in \mathcal{S}_{\lambda}}
\end{equation}
for all $\lambda \in \bar{\Lambda}$ and random variables $\hat{X}$ satisfying
$S \markovlink X \markovlink \hat{X}$.
Note that the right-hand side of \eqref{e:results.properties.classical} is the
classical RD function of $X$ for the distortion measures
$\{ \hat{d}_{\lambda} \}_{\lambda \in \bar{\Lambda}}$.
By the monotonicity and convexity of the right-hand side of
\eqref{e:results.properties.classical} with respect to
$\{ D ( \lambda ) \}_{\lambda \in \Lambda}$, $R^{*}$ is non-increasing and
convex.
If $\mathcal{X}$ and $\hat{\mathcal{X}}$ are finite, we can compute $R^{*}$
numerically using the Blahut-Arimoto algorithm \cite{arimoto1972, blahut1972}.

We can use the total expectation formula to prove the following corollary.

\begin{corollary}
    \label{cor:results.properties.omos} % one measure for one subsource
    Let $\Lambda = \mathcal{S}$, $d$ be a distortion measure on
    $\mathcal{X} \times \hat{\mathcal{X}}$, and $R^{\mathrm{O}}$ be the
    classical RD function of $X$ for $d$.
    Suppose $d_{\lambda} = d$ and $\mathcal{S}_{\lambda} = \{ \lambda \}$ for
    all $\lambda \in \Lambda$.
    Then for every $\delta \in \cointerval{0}{\infty}$ we have
    \begin{equation}
        R^{\mathrm{O}} ( \delta )
        = \inf_{D \in A ( \delta )} R^{*} ( D ) ,
        \label{e:results.properties.omosr} % one-measure-for-one-subsource rate
    \end{equation}
    where $A ( \delta )$ is the set of
    $\{ D ( \lambda ) \}_{\lambda \in \Lambda}$ satisfying
    \begin{equation}
        \sum_{s \in \mathcal{S}} \probability{S = s} D ( s ) \le \delta .
    \end{equation}
\end{corollary}

\begin{figure}[!t]
    \centering
    \includegraphics[scale = 0.4]{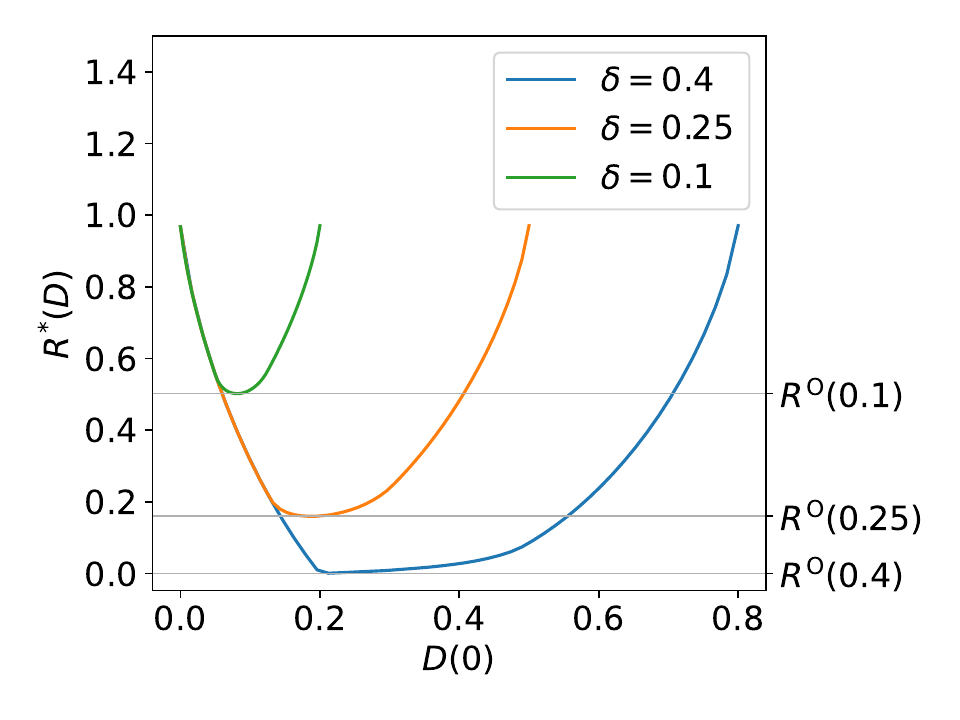}
    \caption{%
        The relation between $R^{\mathrm{O}}$ and $R^{*}$ in
        Example~\ref{exp:results.properties.omos}.
        Each curve shows the relation between $D ( 0 )$ and $R^{*} ( D )$ under
        the constraint
        $\probability{S = 0} D ( 0 ) + \probability{S = 1} D ( 1 ) = \delta$.%
    }
    \label{f:results.properties.omosrates}
\end{figure}

\begin{example}
    \label{exp:results.properties.omos}
    Let
    $\mathcal{S} = \mathcal{X} = \hat{\mathcal{X}} = \Lambda = \{ 0 , 1 \}$,
    \begin{equation}
        \begin{bmatrix}
            \probability{S = 0 , X = 0} & \probability{S = 0 , X = 1} \\
            \probability{S = 1 , X = 0} & \probability{S = 1 , X = 1}
        \end{bmatrix} \!
        = \! \begin{bmatrix}
            0.1 & 0.4 \\
            0.3 & 0.2
        \end{bmatrix} \! ,
    \end{equation}
    $\mathcal{S}_{0} = \{ 0 \}$, $\mathcal{S}_{1} = \{ 1 \}$, and
    $d_{0} = d_{1} = d$, where $d$ is the Hamming distortion measure on
    $\mathcal{X} \times \hat{\mathcal{X}}$ defined by
    \begin{equation}
        d ( x , \hat{x} )
        = \begin{cases}
            0 , & x = \hat{x} , \\
            1 , & x \not= \hat{x} .
        \end{cases}
    \end{equation}
    \figurename~\ref{f:results.properties.omosrates} shows the relation between
    $R^{\mathrm{O}}$ in Corollary~\ref{cor:results.properties.omos} and
    $R^{*}$.
    We can see
    \begin{enumerate}
        \item Subject to
        \begin{equation}
            \probability{S = 0} D ( 0 ) + \probability{S = 1} D ( 1 ) = 0.25 ,
            \label{e:results.proeprties.avdis} % average distortion
        \end{equation}
        $R^{*} ( D )$ is minimized when $D ( 0 )$ is approximately $0.2$;
        \item The minimum $R^{*} ( D )$ subject to
        \eqref{e:results.proeprties.avdis} equals $R^{\mathrm{O}} ( 0.25 )$;
        and
        \item $R^{*} ( D ) > R^{\mathrm{O}} ( 0.25 )$ when
        \eqref{e:results.proeprties.avdis} and $D ( 0 ) = 0.25$ hold, i.e. when
        $D ( 0 ) = D ( 1 ) = 0.25$.
    \end{enumerate}
    Analogous facts can be seen for $\delta = 0.1$ and for $\delta = 0.4$.
    The second observation confirms
    Corollary~\ref{cor:results.properties.omos}.
    The third observation is more interesting.
    If $D ( 0 ) = D ( 1 ) = 0.25$, then $R^{*} ( D )$ is the minimum rate we
    need to ensure that the distortion level of each subsource does not exceed
    $0.25$.
    We need the rate $R^{\mathrm{O}} ( 0.25 )$ to ensure that the average
    distortion of all symbols output by the composite source does not exceed
    $0.25$.
    It is not difficult to understand the fact
    $R^{*} ( D ) > R^{\mathrm{O}} ( 0.25 )$: If the distortion level of each
    subsource does not exceed $0.25$, then the average distortion does not
    exceed $0.25$, but not vice versa.
    Hence, fidelity criteria on individual subsources can be stricter than the
    classical fidelity criteria that the average distortion of all symbols does
    not exceed a certain distortion level.
\end{example}

\begin{corollary}
    \label{cor:results.properties.admeasure} % an additional distortion measure
    Let $\tilde{\Lambda} = \Lambda \cup \{ \tilde{\lambda} \}$, where
    $\tilde{\lambda} \not\in \Lambda$.
    Define distortion measures
    $\{ \tilde{d}_{\lambda} \}_{\lambda \in \tilde{\Lambda}}$ on
    $\mathcal{X} \times ( \hat{\mathcal{X}} \times \mathcal{U} )$ by
    \begin{equation}
        \tilde{d}_{\lambda} ( x , ( \hat{x} , u ) )
        = \begin{cases}
            d_{\lambda} ( x , \hat{x} ) , &
            \lambda \in \Lambda , \\
            0 , &
            \lambda = \tilde{\lambda} , c ( x ) = u , \\
            1 , &
            \lambda = \tilde{\lambda} , c ( x ) \not= u .
        \end{cases}
    \end{equation}
    Let $\tilde{\mathcal{S}}_{\tilde{\lambda}} = \mathcal{S}$ and
    $\tilde{\mathcal{S}}_{\lambda} = \mathcal{S}_{\lambda}$ for all
    $\lambda \in \Lambda$.
    Then for every $\{ D ( \lambda ) \}_{\lambda \in \Lambda}$, we have
    $R^{\mathrm{C}} ( D ) = \tilde{R} ( \tilde{D} )$, where $\tilde{R}$ is the
    RD function of $( S , X )$ for $\{
        ( \tilde{d}_{\lambda} , \tilde{\mathcal{S}}_{\lambda} )
    \}_{\lambda \in \tilde{\Lambda}}$, $\tilde{D} ( \tilde{\lambda} ) = 0$, and
    $\tilde{D} ( \lambda ) = D ( \lambda )$ for all $\lambda \in \Lambda$.
\end{corollary}

Since the distortion measures
$\{ \tilde{d}_{\lambda} \}_{\lambda \in \tilde{\Lambda}}$ in
Corollary~\ref{cor:results.properties.admeasure} are on
$\mathcal{X} \times ( \hat{\mathcal{X}} \times \mathcal{U} )$, the RD function
$\tilde{R}$ is for a situation where the decoder outputs not only symbols
taking values in $\hat{\mathcal{X}}$, but also labels taking values in
$\mathcal{U}$.
For $\lambda \in \Lambda$, the fidelity criterion $(
    \tilde{d}_{\lambda} ,
    \tilde{\mathcal{S}}_{\lambda} ,
    \tilde{D} ( \lambda )
) = ( \tilde{d}_{\lambda} , \mathcal{S}_{\lambda} , D ( \lambda ) )$ ignores
the labels and imposes the same distortion constraint as that imposed by the
fidelity criterion $( d_{\lambda} , \mathcal{S}_{\lambda} , D ( \lambda ) )$.
The distortion measure $\tilde{d}_{\tilde{\lambda}}$ shows if the decoder
reproduces the correct label.
Hence the fidelity criterion $(
    \tilde{d}_{\tilde{\lambda}} ,
    \tilde{\mathcal{S}}_{\tilde{\lambda}} ,
    \tilde{D} ( \tilde{\lambda} )
) = ( \tilde{d}_{\tilde{\lambda}} , \mathcal{S} , 0 )$ requires the decoder to
reproduce all the labels correctly.
Similarly,
\begin{equation}
    \mean{
        \tilde{d}_{\tilde{\lambda}} ( X , ( \hat{X} , U ) )
    |
        S \in \tilde{\mathcal{S}}_{\tilde{\lambda}}
    }
    \le \tilde{D} ( \tilde{\lambda} )
\end{equation}
is equivalent to $\probability{c ( X ) = U} = 1$, and for
$\lambda \in \Lambda$,
\begin{equation}
    \mean{
        \tilde{d}_{\lambda} ( X , ( \hat{X} , U ) )
    |
        S \in \tilde{\mathcal{S}}_{\lambda}
    }
    \le \tilde{D} ( \lambda )
\end{equation}
is equivalent to $
    \mean{d_{\lambda} ( X , \hat{X} ) | S \in \mathcal{S}_{\lambda}}
    \le D ( \lambda )
$.
These lead to Corollary~\ref{cor:results.properties.admeasure}.

We have shown by using Corollary~\ref{cor:results.properties.classical} that
$R^{*}$ is non-increasing and convex.
Similar arguments show that $\tilde{R}$ in
Corollary~\ref{cor:results.properties.admeasure} is non-increasing and convex.
By Corollaries~\ref{cor:results.properties.admeasure} and
\ref{cor:results.properties.order}, $R^{\mathrm{C}}$ is a special case of
$\tilde{R}$ with $\tilde{D} ( \tilde{\lambda} ) = 0$, and
$R^{\mathrm{G}} ( D ) = R^{\mathrm{C}} ( D ) - \entropy{c ( X )}$ for all
$\{ D ( \lambda ) \}_{\lambda \in \Lambda}$.
We thus have the following corollary.

\begin{corollary}
    \label{cor:results.properties.convexity}
    The RD functions $R^{*}$, $R^{\mathrm{C}}$, and $R^{\mathrm{G}}$ are
    non-increasing and convex.
\end{corollary}

The following theorem can be viewed as a generalization of a property of the
conditional RD function \cite{gray1972, gray1973}.

\begin{theorem}
    \label{thm:results.properties.allocation} % 码率和失真的分配
    Define $
        \bar{\mathcal{U}}
        = \{ u \in \mathcal{U} | \probability{c ( X ) = u} > 0 \}
    $.
    For $u \in \bar{\mathcal{U}}$, let $P_{u}$ be the conditional distribution
    of $( S , X )$ given $c ( X ) = u$, and $R_{u}^{*}$ be the RD function
    defined in Section~\ref{u:definitions.rdfunctions} of a $P_{u}$-distributed
    random pair for
    $\{ ( d_{\lambda} , \mathcal{S}_{\lambda} ) \}_{\lambda \in \Lambda}$.
    Then for every $\{ D ( \lambda ) \}_{\lambda \in \Lambda}$,
    \begin{equation}
        R^{\mathrm{G}} ( D )
        = \inf_{\delta \in \Delta ( D )}
        \sum_{u \in \bar{\mathcal{U}}}
        \probability{c ( X ) = u}
        R_{u}^{*} ( \delta ( \cdot , u ) ) ,
        \label{e:results.properties.ratall} % rate allocation
    \end{equation}
    where $\Delta ( D )$ is the set of functions
    $\delta : \Lambda \times \mathcal{U} \to \cointerval{0}{\infty}$ satisfying
    \begin{equation}
        \sum_{u \in \bar{\mathcal{U}}}
        \probability{S \in \mathcal{S}_{\lambda} , c ( X ) = u}
        \delta ( \lambda , u )
        \le \probability{S \in \mathcal{S}_{\lambda}} D ( \lambda )
        \label{e:results.properties.disall} % distortion allocation
    \end{equation}
    for all $\lambda \in \Lambda$.
\end{theorem}

If $\mathcal{S}_{\lambda} = \mathcal{S}$ for all $\lambda \in \Lambda$, then
$R^{\mathrm{G}}$ is a conditional RD function and each $R_{u}^{*}$ in
Theorem~\ref{thm:results.properties.allocation} is a classical RD function.
Hence, \eqref{e:results.properties.ratall} is a generalization of (2.2) in
\cite{gray1973} for subsource-dependent fidelity criteria.
Theorem~\ref{thm:results.properties.allocation} also provides an approach to
numerical computation of $R^{\mathrm{G}}$, because we can compute $R_{u}^{*}$
using the Blahut-Arimoto algorithm, as pointed out in the discussions following
Corollary~\ref{cor:results.properties.classical}.

    % cspell: words losslessly, subsource, subsources

\subsection{CTC Code Where Labels are Compressed by a Variable-Length Code}
\label{u:results.combination}

We have seen in Section~\ref{u:definitions.codes} that CTC codes can have
different implementations.
Our proofs outlined in Section~\ref{u:sketches.ctc} of
Theorems~\ref{thm:results.theorems.ctc} and
\ref{thm:results.theorems.condition} show that CTC codes implemented as in
\figurename~\ref{f:definitions.codes.labelctc} can have the performance
characterized by $R^{\mathrm{C}}$.
The implementation shown in \figurename~\ref{f:definitions.codes.labelctc},
i.e. the implementation where labels are compressed by a variable-length code,
is hence justified.
Now we illustrate this by showing the relevant steps of the proofs.

Define $\bar{\mathcal{U}}$ and $\{ P_{u} \}_{u \in \bar{\mathcal{U}}}$ as in
Theorem~\ref{thm:results.properties.allocation}.
For $u \in \bar{\mathcal{U}}$, let an
$( \mathcal{X} , \hat{\mathcal{X}} )$-variable-length code
$( f_{u} , \varphi_{u} )$ achieves $( R_{u} , \delta ( \cdot , u ) )$ for a
$P_{u}$-distributed random pair and
$\{ ( d_{\lambda} , \mathcal{S}_{\lambda} ) \}_{\lambda \in \Lambda}$.
Here $\delta ( \cdot , u )$ denotes distortion levels
$\{ \delta ( \lambda , u ) \}_{\lambda \in \Lambda}$.
Then we can combine the codes
$\{ ( f_{u} , \varphi_{u} ) \}_{u \in \bar{\mathcal{U}}}$ and a variable-length
code that losslessly compresses class labels in the way shown in
\figurename~\ref{f:definitions.codes.labelctc} to get an
$( \mathcal{X} , \hat{\mathcal{X}} , \mathcal{U} , c )$-CTC code
$( f , \varphi )$.
By Lemmas~\ref{lem:sketches.ctc.distribution},
\ref{lem:sketches.ctc.asymlength}, and \ref{lem:sketches.ctc.fidelity} stated
in Section~\ref{u:sketches.ctc}, the rate of $( f , \varphi )$ is approximately
\begin{equation}
    \entropy{c ( X )}
    + \sum_{u \in \bar{\mathcal{U}}} \probability{c ( X ) = u} R_{u} ,
    \label{e:results.combination.rate}
\end{equation}
and for $\lambda \in \bar{\Lambda}$, the distortion level of $( f , \varphi )$
corresponding to $d_{\lambda}$ and $\mathcal{S}_{\lambda}$ is approximately
\begin{equation}
    \sum_{u \in \bar{\mathcal{U}}}
    \probability{c ( X ) = u | S \in \mathcal{S}_{\lambda}}
    \delta ( \lambda , u ) .
    \label{e:results.combination.distortion}
\end{equation}

For a family $\{ D ( \lambda ) \}_{\lambda \in \Lambda}$ of distortion levels,
we should choose such $\delta$ and $\{ R_{u} \}_{u \in \bar{\mathcal{U}}}$ that
the approximate distortion level \eqref{e:results.combination.distortion} does
not exceed $D ( \lambda )$ for any $\lambda$ and that
$R_{u} = R_{u}^{*} ( \delta ( \cdot , u ) )$ for all $u \in \bar{\mathcal{U}}$.
By $R^{\mathrm{C}} ( D ) = R^{\mathrm{G}} ( D ) + \entropy{c ( X )}$ and
Theorem~\ref{thm:results.properties.allocation}, the minimum rate
\eqref{e:results.combination.rate} for $\delta$ so chosen is
$R^{\mathrm{C}} ( D )$.
Hence CTC codes where labels are compressed by a variable-length code can have
the performance characterized by $R^{\mathrm{C}}$.
The performance of other implementations of CTC codes is left for future
research.

    % cspell: words subsources

\subsection{A Case with Perfect Classification}
\label{u:results.case}

\begin{figure}[!t]
    \centering
    \includegraphics[scale = 0.4]{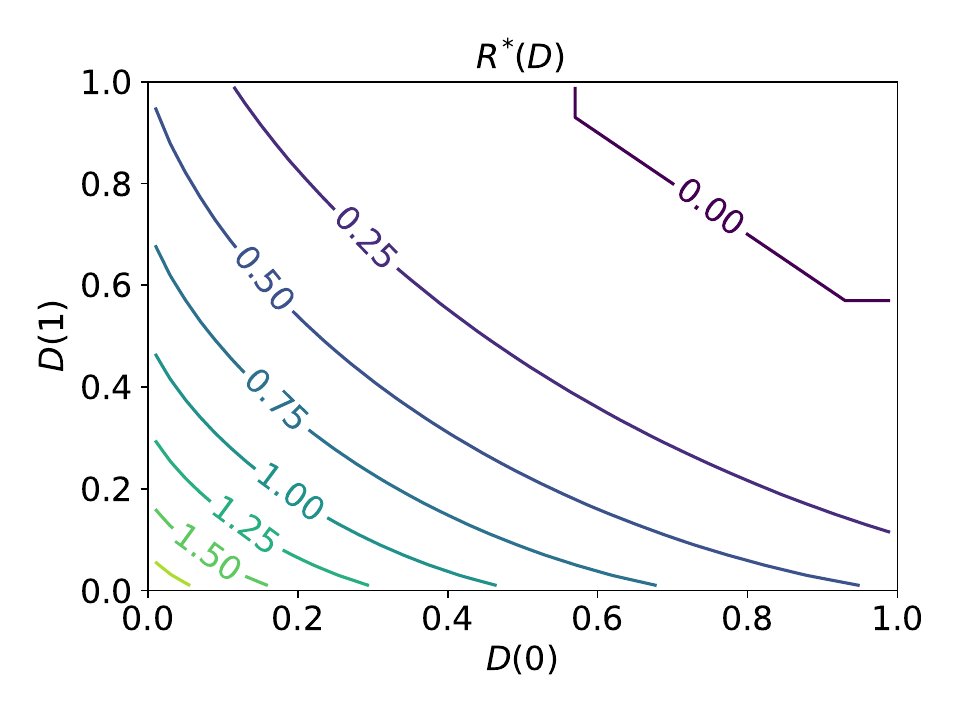}
    \caption{%
        The RD function $R^{*}$ in Example~\ref{exp:results.case.hamming}.}
    \label{f:results.case.rdf}
\end{figure}

\begin{figure}[!t]
    \centering
    \includegraphics[scale = 0.4]{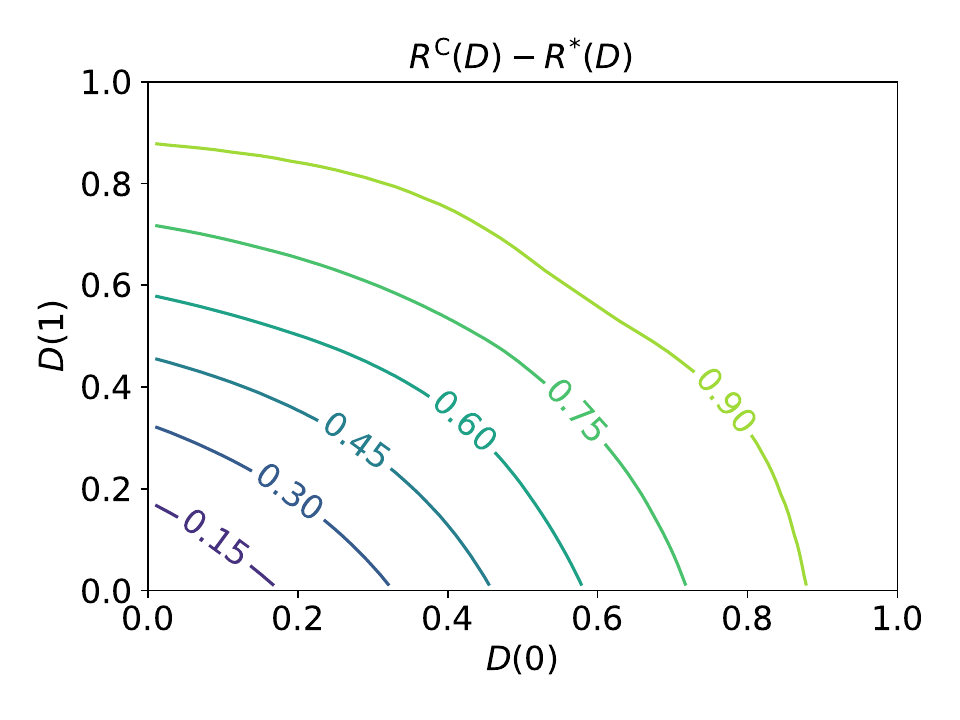}
    \caption{%
        The gap $R^{\mathrm{C}} - R^{*}$ in
        Example~\ref{exp:results.case.hamming}.}
    \label{f:results.case.gap}
\end{figure}

In this section, we study a case where the classification is perfect in that
$S \in \mathcal{S}_{\lambda}$ is equivalent to $c ( X ) = \lambda$ with
probability one for all $\lambda \in \Lambda$.
Of course, this is possible only if all $\lambda \in \Lambda$ satisfying
$\probability{S \in \mathcal{S}_{\lambda}} > 0$ are chosen as labels.
Insights into connections between CTC codes and general variable-length codes
will be gained in the following theorem and example.

\begin{theorem}
    \label{thm:results.case.case}
    For $\lambda \in \bar{\Lambda}$, let $Q_{\lambda}$ be the distribution of
    $X$ given the condition $S \in \mathcal{S}_{\lambda}$ and $R^{( \lambda )}$
    be the RD function of a $Q_{\lambda}$-distributed random variable for
    $d_{\lambda}$.
    Fix a family $\{ D ( \lambda ) \}_{\lambda \in \Lambda}$ of
    distortion levels.
    If $S \in \mathcal{S}_{\lambda}$ is equivalent to $c ( X ) = \lambda$ with
    probability one for all $\lambda \in \Lambda$, then
    \begin{equation}
        R^{\mathrm{G}} ( D )
        = \sum_{\lambda \in \bar{\Lambda}}
        \probability{S \in \mathcal{S}_{\lambda}}
        R^{( \lambda )} ( D ( \lambda ) ) .
        \label{e:results.case.conrate} % conditional rate
    \end{equation}
    If we have $| \bar{\Lambda} | = 1$ in addition, then
    $R^{\mathrm{G}} ( D ) = R^{*} ( D )$.
\end{theorem}

The meaning of \eqref{e:results.case.conrate} is as follows.
For $\lambda \in \bar{\Lambda}$, the symbols from the subsources in
$\mathcal{S}_{\lambda}$ account for approximately
$\probability{S \in \mathcal{S}_{\lambda}}$ of all the source symbols, and
consequently we need a rate of
$\probability{S \in \mathcal{S}_{\lambda}} R^{( \lambda )} ( D ( \lambda ) )$
to describe them.

\begin{example}
    \label{exp:results.case.hamming}
    Let $\mathcal{S} = \mathcal{X} = \hat{\mathcal{X}} = \{ 0 , 1 , 2 , 3 \}$,
    $S$ be uniformly distributed over $\mathcal{S}$, $S = X$ with probability
    one, $\Lambda = \mathcal{U} = \{ 0 , 1 \}$,
    $\mathcal{S}_{0} = \{ 0 , 1 \}$, $\mathcal{S}_{1} = \{ 2 , 3 \}$,
    $c ( 0 ) = c ( 1 ) = 0$, $c ( 2 ) = c ( 3 ) = 1$, and
    \begin{equation}
        d_{0} ( x , \hat{x} )
        = d_{1} ( x , \hat{x} )
        = \begin{cases}
            0 , & x = \hat{x} , \\
            1 , & x \not= \hat{x} , \\
        \end{cases}
    \end{equation}
    for all $x \in \mathcal{X}$ and $\hat{x} \in \hat{\mathcal{X}}$.
    The RD function $R^{*}$ and the gap $R^{\mathrm{C}} - R^{*}$ are shown in
    \figuresname~\ref{f:results.case.rdf} and \ref{f:results.case.gap},
    respectively.
\end{example}

From \eqref{e:results.properties.igai}, \eqref{e:results.properties.igae}, and
Example~\ref{exp:results.case.hamming} we can see
$R^{\mathrm{G}} ( D ) < R^{*} ( D ) < R^{\mathrm{C}} ( D )$ unless certain
conditions are satisfied.
This is true even if the classification is perfect.
The strict inequality $R^{\mathrm{G}} ( D ) < R^{*} ( D )$ can be understood by
observing that the encoder not only needs to describe source symbols in each
class to the decoder, but also needs to inform the decoder the labels of the
symbols.
On the other hand,
$R^{*} ( D ) < R^{\mathrm{C}} ( D ) = R^{\mathrm{G}} ( D ) + \entropy{c ( X )}$
show that compressing the labels using a lossless code with the rate
$\entropy{c ( X )}$ is usually a suboptimal way to do this.
In the extreme case of $| \bar{\Lambda} | = 1$, $R^{*} ( D )$ even equals
$R^{\mathrm{G}} ( D )$.
The differences between the reproduced symbols from different classes may be
utilized to reduce the additional rate required by label compression.
In conclusion, CTC coding can be suboptimal even if the classification is
perfect.

    % cspell: words losslessly, quantizer, subsource

\subsection{Comparison of RD Functions}
\label{u:results.comparison}

In this section, we will show that the performance loss
$R^{\mathrm{C}} ( D ) - R^{*} ( D )$ is small under mild conditions.
Equation~\eqref{e:results.properties.igae} leads to the following corollary.

\begin{corollary}
    \label{cor:results.comparison.limit}
    Fix $\mu \in \Lambda$ and a family
    $\{ D ( \lambda ) \}_{\lambda \in \Lambda}$ of distortion levels.
    Let $D ( \mu ) > 0$.
    For $\delta \in \cointerval{0}{\infty}$ and $\lambda \in \Lambda$, let
    \begin{equation}
        D_{\delta} ( \lambda )
        = \begin{cases}
            \delta , & \lambda = \mu , \\
            D ( \lambda ) , & \lambda \not= \mu .
        \end{cases}
    \end{equation}
    If there exists a family
    $\{ \hat{X}_{\delta} \}_{\delta \in \oointerval{0}{D ( \mu )}}$ of random
    variables such that
    \begin{enumerate}
        \item $\hat{X}_{\delta} \in \Xi ( D_{\delta} )$ for all
        $\delta \in \oointerval{0}{D ( \mu )}$; and
        \item $\minfo{X ; \hat{X}_{\delta}} - R^{*} ( D_{\delta} )$ and
        $\entropy{c ( X ) | \hat{X}_{\delta}}$ tend to 0 as
        $\oointerval{0}{D ( \mu )} \owns \delta \to 0$;
    \end{enumerate}
    then $R^{\mathrm{C}} ( D_{\delta} ) - R^{*} ( D_{\delta} ) \to 0$ as
    $\oointerval{0}{D ( \mu )} \owns \delta \to 0$.
\end{corollary}

As $\delta \to 0$, $D_{\delta} ( \mu )$ tends to zero while the other
distortion levels in $\{ D_{\delta} ( \lambda ) \}_{\lambda \in \Lambda}$
remain fixed.
Hence under the conditions in Corollary~\ref{cor:results.comparison.limit}, the
performance loss $R^{\mathrm{C}} - R^{*}$ is negligible for asymptotically
small distortion.

The cases where the conditions in Corollary~\ref{cor:results.comparison.limit}
are satisfied are not rare.
For example, we can let $\mathcal{X} = \hat{\mathcal{X}}$ and let $d_{\mu}$ be
a distortion measure that characterizes the closeness or similarity between
source symbols and their reproductions, such as the Hamming distortion measure
or the squared-error distortion measure.
For such a $d_{\mu}$ and some common classifier $c$, a sufficiently small
$d_{\mu} ( x , \hat{x} )$ means that $x$ and $\hat{x}$ tend to have the same
label.
For $\delta \in \oointerval{0}{D ( \mu )}$, we let $\hat{X}_{\delta}$ be a
random variable in $\Xi ( D_{\delta} )$ satisfying
$\minfo{X ; \hat{X}_{\delta}} < R^{*} ( D_{\delta} ) + \delta$.
Further assume that the support set of $X$ equals the conditional support set
of $X$ given $S \in \mathcal{S}_{\mu}$.
Then for sufficiently small $\delta$, $\mean{d_{\mu} ( X , \hat{X}_{\delta} )}$
is small, and consequently $c ( X ) = c ( \hat{X}_{\delta} )$ with a high
probability.
Hence for sufficiently small $\delta$, $\entropy{c ( X ) | \hat{X}_{\delta}}$
is small, and by Corollary~\ref{cor:results.comparison.limit}, we can expect
$R^{\mathrm{C}} ( D_{\delta} ) - R^{*} ( D_{\delta} )$ to be small.

If $D ( \lambda )$ is sufficiently large for all $\lambda \in \Lambda$, then
$R^{\mathrm{G}} ( D ) = R^{*} ( D ) = 0$.
Therefore, $R^{*} ( D )$ can be close to $R^{\mathrm{G}} ( D )$ for $D$ having
large values and can be close to $R^{\mathrm{C}} ( D )$ for $D$ having small
values.
Or equivalently, the performance loss $R^{\mathrm{C}} ( D ) - R^{*} ( D )$ of
optimal CTC codes is close to $0$ for $D$ having small values and is close to
$\entropy{c ( X )}$ for $D$ having large values.
This can be understood by observing that lossless compression of labels needs a
rate of $\entropy{c ( X )}$, and that CTC codes need to compress the labels
losslessly even if $D ( \lambda )$ is large.
For $D$ having medium values, we have
$0 < R^{\mathrm{C}} ( D ) - R^{*} ( D ) < \entropy{c ( X )}$ by the continuity
of $R^{*}$ and $R^{\mathrm{C}}$, which follows from
Corollary~\ref{cor:results.properties.convexity}.
The performance loss $R^{\mathrm{C}} ( D ) - R^{*} ( D )$ is small if the
number $| \mathcal{U} |$ of classes is small.

\begin{figure}[!t]
    \centering
    \includegraphics[scale = 0.4]{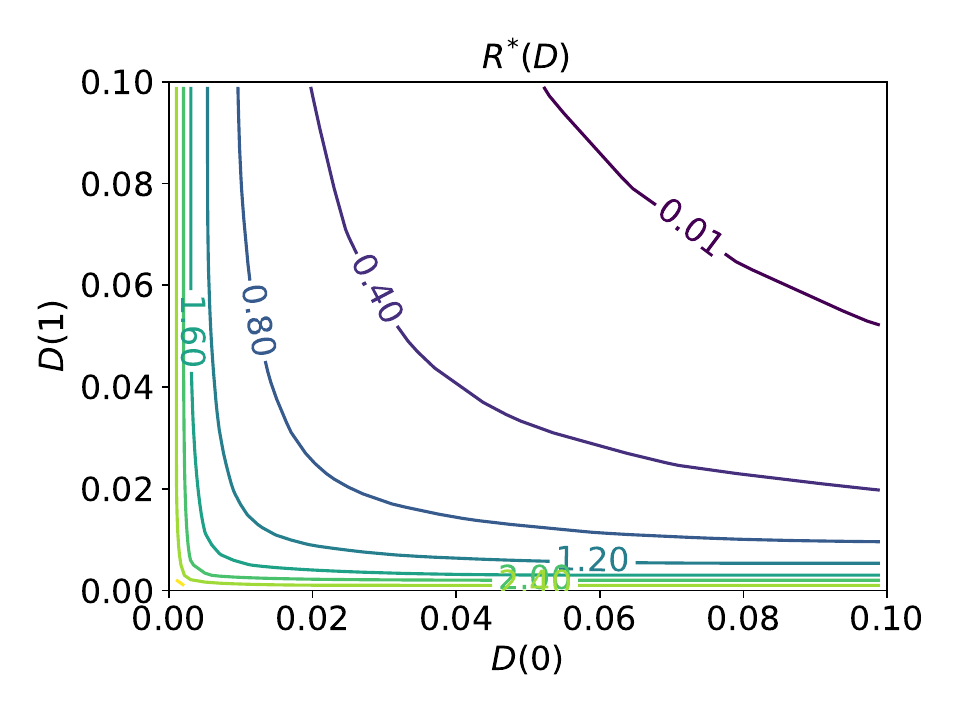}
    \caption{%
        The RD function $R^{*}$ in Example~\ref{exp:results.comparison.gmm}.}
    \label{f:results.comparison.rdf}
\end{figure}

\begin{figure}[!t]
    \centering
    \includegraphics[scale = 0.4]{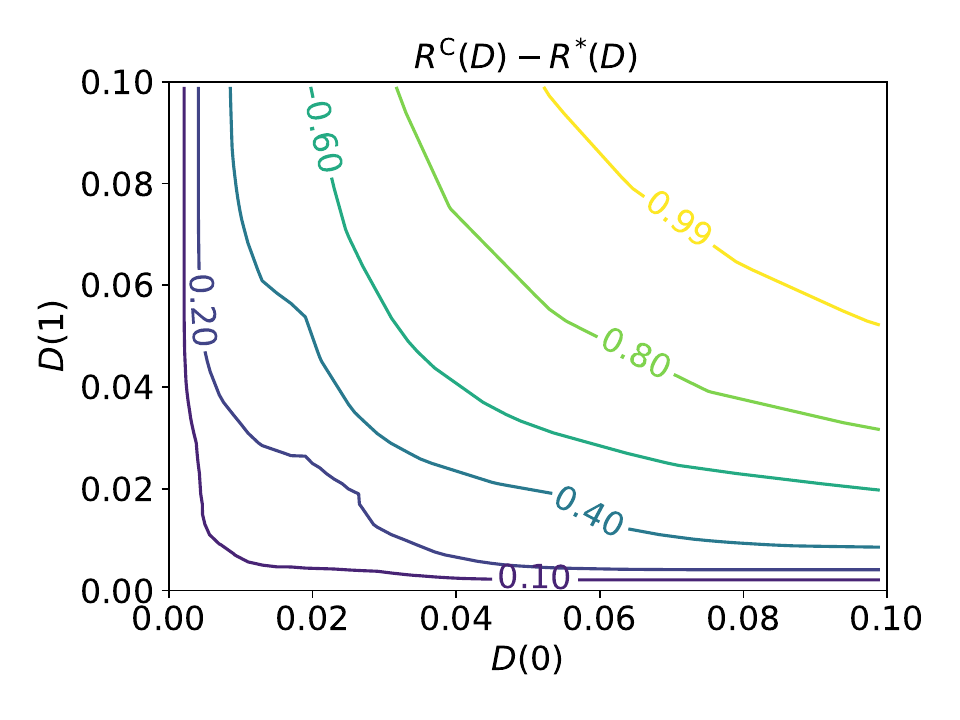}
    \caption{%
        The gap $R^{\mathrm{C}} - R^{*}$ in
        Example~\ref{exp:results.comparison.gmm}.}
    \label{f:results.comparison.cgap}
\end{figure}

\begin{example}
    \label{exp:results.comparison.gmm} % Gaussian mixture model
    Let $S$ be uniformly distributed over $\mathcal{S} = \{ 0 , 1 \}$, $
        \mathcal{X}
        = \hat{\mathcal{X}}
        = \{ n / 256 + 1 / 512 | n \in \conumbers{256} \}
    $, $\Lambda = \mathcal{U} = \{ 0 , 1 \}$, $\mathcal{S}_{0} = \{ 0 \}$,
    $\mathcal{S}_{1} = \{ 1 \}$, and
    $d_{0} ( x , \hat{x} ) = d_{1} ( x , \hat{x} ) = ( x - \hat{x} )^{2}$ for
    all $x \in \mathcal{X}$ and $\hat{x} \in \hat{\mathcal{X}}$.
    Define an 8-bit quantizer $Q : \reals \to \mathcal{X}$ by
    \begin{equation}
        Q ( y )
        = \begin{cases}
            1 / 512 ,
            & y < 1 / 256 , \\
            \dfrac{\lfloor 256 y \rfloor}{256} + \dfrac{1}{512} , &
            \dfrac{1}{256} \le y < \dfrac{255}{256} , \\
            511 / 512 , &
            255 / 256 \le y .
        \end{cases}
    \end{equation}
    Let a random variable $Y$ have the conditional distribution
    $\dnormal{0.3}{0.04}$ given $S = 0$, and have the conditional distribution
    $\dnormal{0.7}{0.04}$ given $S = 1$.
    Suppose $X = Q ( Y )$ with probability one.
    A reasonable classifier $c$ assigns label $0$ to every $x < 1 / 2$, and
    assigns label $1$ to every $x \ge 1 / 2$.
    The RD function $R^{*}$ is shown in
    \figurename~\ref{f:results.comparison.rdf}.
    \figurename~\ref{f:results.comparison.cgap} shows the gap between $R^{*}$
    and $R^{\mathrm{C}}$.
    The performance loss $R^{\mathrm{C}} ( D ) - R^{*} ( D )$ is close to
    $1 = \entropy{c ( X )}$ when $D ( 0 )$ and $D ( 1 )$ are sufficiently large
    and is close to 0 when $D ( 0 )$ or $D ( 1 )$ are close to 0.
    This confirms Corollary~\ref{cor:results.comparison.limit} and the
    discussions following it.
\end{example}

    \section{Proof Sketches of Coding Theorems}
    \label{s:sketches}
    \subsection{Theorem~\ref{thm:results.theorems.general}}
\label{u:sketches.general}

A key step in the proof is to define functions
$\{ d_{\lambda , \delta} \}_{\lambda \in \Lambda , \delta \in \reals}$ from
$\mathcal{X} \times \hat{\mathcal{X}} \times \mathcal{S}$ to $\reals$ by
\begin{equation}
    d_{\lambda , \delta} ( x , \hat{x} , s )
    = \begin{cases}
        d_{\lambda} ( x , \hat{x} ) , &
        s \in \mathcal{S}_{\lambda} , \\
        \delta , &
        s \not\in \mathcal{S}_{\lambda} .
    \end{cases}
    \label{e:sketches.general.dismeas} % distortion measure
\end{equation}
For $\lambda \in \Lambda$ and $\delta \in \cointerval{0}{\infty}$,
$d_{\lambda , \delta}$ can be interpreted as a distortion measure on
$\mathcal{X} \times \hat{\mathcal{X}}$ dependent on $s$.
For $\lambda \in \Lambda$, $\delta \in \cointerval{0}{\infty}$,
$n \in \cointegers{1}{\infty}$, $\{ s_{i} \}_{i = 1}^{n} \in \mathcal{S}^{n}$,
$\{ x_{i} \}_{i = 1}^{n} \in \mathcal{X}^{n}$, and
$\{ \hat{x}_{i} \}_{i = 1}^{n} \in \hat{\mathcal{X}}^{n}$, we can verify that
\begin{equation}
    \frac{1}{n}
    \sum_{i = 1}^{n}
    d_{\lambda , \delta} ( x_{i} , \hat{x}_{i} , s_{i} )
    \le \delta
    \label{e:sketches.general.avdis} % average distortion
\end{equation}
is a sufficient and necessary condition for $s_{\ocnumbers{n}}$,
$x_{\ocnumbers{n}}$, and $\hat{x}_{\ocnumbers{n}}$ to meet the fidelity
criterion $( d_{\lambda} , \mathcal{S}_{\lambda} , \delta )$.
The inequality \eqref{e:sketches.general.avdis} means that the average of all
$d_{\lambda} ( x_{i} , \hat{x}_{i} )$ satisfying
$s_{i} \in \mathcal{S}_{\lambda}$ does not exceed $\delta$, because
$d_{\lambda , \delta} ( x_{i} , \hat{x}_{i} , s_{i} ) = \delta$ whenever
$s_{i} \not\in \mathcal{S}_{\lambda}$.
Thus, the problem is reduced to an RD problem where the distortion measures are
influenced by some side information unknown to both the encoder and the decoder
\cite{linder2000, martinian2008}.
In spite of the difference that the distortion measures depend on the
distortion level $\delta$, Theorem~\ref{thm:results.theorems.general} can be
proved using existing results \cite{martinian2008} and the following corollary
of the definition of $\Xi ( D )$.

\begin{corollary}
    \label{cor:sketches.general.rdfunction} % rate-distortion function
    Define $
        d_{\lambda , \delta} :
        \mathcal{X} \times \hat{\mathcal{X}} \times \mathcal{S} \to \reals
    $ by \eqref{e:sketches.general.dismeas} for $\lambda \in \Lambda$ and
    $\delta \in \reals$.
    Let $\{ D ( \lambda ) \}_{\lambda \in \Lambda}$ be a family of distortion
    levels.
    Then $\Xi ( D )$ equals the set of random variables $\hat{X}$ satisfying
    $S \markovlink X \markovlink \hat{X}$ and
    $\mean{d_{\lambda , D ( \lambda )} ( X , \hat{X} , S )} \le D ( \lambda )$
    for all $\lambda \in \Lambda$.
\end{corollary}

Since the existing results are on block codes (see
Section~\ref{u:definitions.codes}), we still need to adapt them for
variable-length codes.
From a sequence of block codes, a variable-length code with almost the same
rate can be constructed using the compound representation of positive integers
\cite{elias1975}.
The compound representation of $j \in \cointegers{1}{\infty}$ is the bit string
$\gamma ( j )$ obtained by
\begin{enumerate}
    \item Writing $j$ in the binary form
    $1 b_{\lfloor \log_{2} ( j ) \rfloor - 1} \cdots b_{1} b_{0}$;
    \item Inserting a $1$ after $b_{0}$ and $0$'s after all the other bits in
    the binary form; and
    \item Dropping the leading $1$ to get the bit string $\gamma ( j )$.
\end{enumerate}
For example, the binary form of 5 is 101, and so $\gamma ( 5 ) = 00011$.
It is known \cite{elias1975} that
\begin{equation}
    | \gamma ( j ) |
    = 1 + 2 \lfloor \log_{2} ( j ) \rfloor
    \le 1 + 2 \log_{2} ( j )
    \label{e:sketches.general.length_cr}
    % ``length_cr'' 表示 ``length of the compound representation''.
\end{equation}
for $j \in \cointegers{1}{\infty}$, and that $\gamma$ is prefix-free, that is,
$\gamma ( i )$ is not a prefix of $\gamma ( j )$ for all $i$,
$j \in \cointegers{1}{\infty}$ when $i \not= j$.
The following lemma shows properties of so constructed variable-length codes.

\begin{lemma}
    \label{lem:sketches.general.vlcc} % variable-length code construction
    For $n \in \cointegers{1}{\infty}$, let $M_{n} \in \cointegers{1}{\infty}$,
    $f_{n}$ be a function from $\mathcal{X}^{n}$ to $\ocnumbers{M_{n}}$, and
    $\varphi_{n}$ be a function from $\ocnumbers{M_{n}}$ to
    $\hat{\mathcal{X}}^{n}$.
    If $R \in \reals$ and $M_{n} \le 2^{n R}$ for all
    $n \in \cointegers{1}{\infty}$, then there exists an
    $( \mathcal{X} , \hat{\mathcal{X}} )$-variable-length code
    $( f , \varphi )$ satisfying
    \begin{equation}
        \limsup_{n \to \infty}
        \frac{1}{n}
        \max_{\boldsymbol{x} \in \mathcal{X}^{n}}
        | f ( \boldsymbol{x} ) |
        \le R \label{e:sketches.general.rate}
    \end{equation}
    and $\varphi ( f ( \boldsymbol{x} ) ) = \varphi_{| \boldsymbol{x} |} (
        f_{| \boldsymbol{x} |} ( \boldsymbol{x} )
    )$ for all non-empty strings $\boldsymbol{x} \in \mathcal{X}^{*}$.
\end{lemma}

The complete proof of Theorem~\ref{thm:results.theorems.general} is given in
Appendix~\ref{s:genproof}.

    % cspell: words losslessly, subsource

\subsection{%
    Theorems~\ref{thm:results.theorems.ctc} and
    \ref{thm:results.theorems.condition}}
\label{u:sketches.ctc}

We prove the converse part of Theorem~\ref{thm:results.theorems.ctc} using
Theorem~\ref{thm:results.theorems.general} and
Corollary~\ref{cor:results.properties.admeasure}.
By assumption, the labels of the source symbols can be correctly recovered from
the transmitted bit string $f ( X_{\ocnumbers{n}} )$ in the code
$( f , \varphi )$.
Hence we can use $( f , \varphi )$ to construct an
$( \mathcal{X} , \hat{\mathcal{X}} \times \mathcal{U} )$-variable-length code
$( f , \tilde{\varphi} )$ that achieves $( R , \tilde{D} )$ for $( S , X )$ and
$\{
    ( \tilde{d}_{\lambda} , \tilde{\mathcal{S}}_{\lambda} )
\}_{\lambda \in \tilde{\Lambda}}$, where $\tilde{D}$, $\tilde{d}_{\lambda}$,
and $\tilde{\mathcal{S}}_{\lambda}$ are defined in
Corollary~\ref{cor:results.properties.admeasure}.
By Theorem~\ref{thm:results.theorems.general} we have
$R \ge \tilde{R} ( \tilde{D} )$, i.e. $R \ge R^{\mathrm{C}} ( D )$.

It is not difficult to prove the direct part of
Theorem~\ref{thm:results.theorems.condition}.
We have shown in Section~\ref{u:sketches.general} how the subsource-dependent
fidelity criteria can be converted to fidelity criteria considered in
\cite{linder2000} and \cite{martinian2008}.
The conversion can be used here to reduce the problem to an RD problem with the
side information expressed as two random variables: $S$ unknown to both the
encoder and the decoder, and $c ( X )$ known by both the encoder and the
decoder.
Then we can use existing results \cite{gray1972, martinian2008} to show that
the lowest achievable rate is $R^{\mathrm{G}} ( D )$.

We will go into more detail about an alternative proof of the direct part of
Theorem~\ref{thm:results.theorems.condition} that also serves as a foundation of
the analysis in Section~\ref{u:results.combination}.
This proof explicitly constructs a variable-length code for each class
$u \in \mathcal{U}$ and then combines the variable-length codes to form the
desired label-based code.
The proof uses the following lemmas.

\begin{lemma}
    \label{lem:sketches.ctc.distribution}
    Let $n \in \cointegers{1}{\infty}$, $u \in \mathcal{U}$, $P_{u}$ be the
    conditional distribution of $( S , X )$ given $c ( X ) = u$, and $J$ be the
    random set $\{ i \in \ocnumbers{n} | c ( X_{i} ) = u \}$.
    Define random pairs $\{ ( T_{i} , Y_{i} ) \}_{i = 1}^{\infty}$ such that
    \begin{enumerate}
        \item For every $i \in \cointegers{1}{\infty}$, $( T_{i} , Y_{i} )$ has
        the distribution $P_{u}$ and is independent of $c ( X_{i} )$; and
        \item $\{ ( T_{i} , Y_{i} , c ( X_{i} ) ) \}_{i = 1}^{\infty}$ is an
        i.i.d. process.
    \end{enumerate}
    Then $( S_{J} , X_{J} )$ and $( T_{J} , Y_{J} )$ are identically
    distributed.
\end{lemma}

\begin{lemma}
    \label{lem:sketches.ctc.asymlength} % variable-length blocks
    Let $\{ Y_{i} \}_{i = 1}^{\infty}$ be a sequence of i.i.d. random variables
    taking values in a set $\mathcal{Y}$.
    For $n \in \cointegers{1}{\infty}$, let $J ( n )$ be a random subset of a
    finite subset of $\cointegers{1}{\infty}$ and let $J ( n )$ be independent
    of $\{ Y_{i} \}_{i = 1}^{\infty}$.
    If $\lim_{n \to \infty} \probability{| J ( n ) | > k} = 1$ for all
    $k \in \cointegers{1}{\infty}$, $G \subseteq \mathcal{Y}^{*}$, and
    $\lim_{k \to \infty} \probability{Y_{\ocnumbers{k}} \in G} = 1$, then
    $\lim_{n \to \infty} \probability{Y_{J ( n )} \in G} = 1$.
\end{lemma}

\begin{lemma}
    \label{lem:sketches.ctc.fidelity}
    Let $d$ be a distortion measure on $\mathcal{X} \times \hat{\mathcal{X}}$,
    $\mathcal{S}_{*}$ be a subset of $\mathcal{S}$,
    $\delta \in \cointerval{0}{\infty}$,
    $\{ s_{i} \}_{i = 1}^{n} \in \mathcal{S}^{n}$,
    $\{ x_{i} \}_{i = 1}^{n} \in \mathcal{X}^{n}$,
    $\{ \hat{x}_{i} \}_{i = 1}^{n} \in \hat{\mathcal{X}}^{n}$,
    $J_{*} = \{ i \in \ocnumbers{n} | s_{i} \in \mathcal{S}_{*} \}$, and
    $\{ J ( u ) \}_{u \in \mathcal{U}}$ be a partition of $\ocnumbers{n}$.
    If $s_{J ( u )}$, $x_{J ( u )}$, and $\hat{x}_{J ( u )}$ meet the fidelity
    criterion $( d , \mathcal{S}_{*} , \delta_{u} )$ for all
    $u \in \mathcal{U}$, and $
        \sum_{u \in \mathcal{U}} | J ( u ) \cap J_{*} | \delta_{u}
        \le | J_{*} | \delta
    $, then $s_{\ocnumbers{n}}$, $x_{\ocnumbers{n}}$, and
    $\hat{x}_{\ocnumbers{n}}$ meet the fidelity criterion
    $( d , \mathcal{S}_{*} , \delta )$.
    If $s_{J ( u )}$, $x_{J ( u )}$, and $\hat{x}_{J ( u )}$ do not meet the
    fidelity criterion $( d , \mathcal{S}_{*} , \delta_{u} )$ for any
    $u \in \mathcal{U}$, and $
        \sum_{u \in \mathcal{U}} | J ( u ) \cap J_{*} | \delta_{u}
        \ge | J_{*} | \delta
    $, then $s_{\ocnumbers{n}}$, $x_{\ocnumbers{n}}$, and
    $\hat{x}_{\ocnumbers{n}}$ do not meet the fidelity criterion
    $( d , \mathcal{S}_{*} , \delta )$.
\end{lemma}

We omit the proofs of the lemmas because they are straightforward.
Lemma~\ref{lem:sketches.ctc.distribution} characterizes the distributions of
the symbols in the classes.
Lemma~\ref{lem:sketches.ctc.asymlength} helps us deal with the nuance that the
number of the source symbols in a class is random.
We use Lemma~\ref{lem:sketches.ctc.fidelity} to calculate the distortion levels
$\{ D ( \lambda ) \}_{\lambda \in \Lambda}$ of the constructed label-based
code.
The direct part of Theorem~\ref{thm:results.theorems.condition} is proved by
showing that the rate of the label-based code is close to
$R^{\mathrm{G}} ( D )$.

Finally, the converse part of Theorem~\ref{thm:results.theorems.condition}
follows from that of Theorem~\ref{thm:results.theorems.ctc}, and the direct
part of Theorem~\ref{thm:results.theorems.ctc} follows from that of
Theorem~\ref{thm:results.theorems.condition}.
The reason is that we can get an
$( \mathcal{X} , \hat{\mathcal{X}} , \mathcal{U} , c )$-CTC code by assembling
the label-based code $( f , \varphi )$ in the direct part of
Theorem~\ref{thm:results.theorems.condition} and a code that compresses class
labels losslessly.
Formally, we have the following lemma.

\begin{lemma}
    \label{lem:sketches.ctc.assemble}
    If $( f_{1} , \varphi_{1} )$ is an
    $( \mathcal{X} , \hat{\mathcal{X}} , \mathcal{U} )$-label-based code,
    $R_{1} \in \cointerval{0}{\infty}$, and
    \begin{equation}
        \lim_{n \to \infty} \probability{
            | f_{1} ( X_{\ocnumbers{n}} , U_{\ocnumbers{n}} ) |
            \le n R_{1}
        }
        = 1 ,
        \label{e:sketches.ctc.parate} % parallel rate
    \end{equation}
    then there exist an $( \mathcal{X} , \hat{\mathcal{X}} )$-variable-length
    code $( f_{2} , \varphi_{2} )$ and a function
    $\psi : \{ 0 , 1 \}^{*} \to \mathcal{U}^{*}$ such that
    \begin{enumerate}
        \item For every $R_{2} > R_{1} + \entropy{c ( X )}$,
        \begin{equation}
            \lim_{n \to \infty}
            \probability{| f_{2} ( X_{\ocnumbers{n}} ) | \le n R_{2}}
            = 1 ; \label{e:sketches.ctc.ctcrate}
        \end{equation}
        and
        \item If $n \in \conumbers{\infty}$,
        $\{ x_{i} \}_{i = 1}^{n} \in \mathcal{X}^{n}$, and
        $u_{i} = c ( x_{i} )$ for all $i \in \ocnumbers{n}$, then
        $\psi ( f_{2} ( x_{\ocnumbers{n}} ) ) = u_{\ocnumbers{n}}$ and
        \begin{equation}
            \varphi_{2} ( f_{2} ( x_{\ocnumbers{n}} ) )
            = \varphi_{1} (
                f_{1} ( x_{\ocnumbers{n}} , u_{\ocnumbers{n}} ) ,
                u_{\ocnumbers{n}}
            ) .
            \label{e:sketches.ctc.idrep} % identical reproduction
        \end{equation}
    \end{enumerate}
\end{lemma}

The complete proofs of Theorems~\ref{thm:results.theorems.ctc} and
\ref{thm:results.theorems.condition} are given in Appendix~\ref{s:ctcproofs}.

    % cspell: words subsource

\section{Conclusion}
\label{s:conclusion}

In this paper, we have proposed subsource-dependent fidelity criteria for
composite sources and solved RD problems based on these criteria.
The criteria and hence the results in the paper are useful in situations where
some data from an information source should be encoded more accurately than
other data from the source.
CTC coding is widely used in such situations.
It is found in this paper that CTC coding in general has performance loss
relative to RD-optimal coding, even if the classification is perfect.
On the other hand, the performance loss is small for appropriately designed CTC
codes.
How to design practical CTC codes is left for future research.

    \appendices

    \section{Proof of Theorem~\ref{thm:results.properties.allocation}}
\label{s:allocation}

We can verify the following lemmas about Markov chains, and will use them in
the proof.

\begin{lemma}
    \label{lem:allocation.condition}
    Suppose $V$, $Y$, and $Z$ are random variables, $B$ is a set, and
    $\probability{Y \in B} > 0$.
    Let $( \tilde{V} , \tilde{Y} , \tilde{Z} ) \sim \tilde{P}$, where
    $\tilde{P}$ is the conditional distribution of $( V , Y , Z )$ given
    $Y \in B$.
    If $V \markovlink Y \markovlink Z$, then
    $\tilde{V} \markovlink \tilde{Y} \markovlink \tilde{Z}$.
\end{lemma}

\begin{lemma}
    \label{lem:allocation.mixture}
    Let $\mathcal{Q}$ be a finite set, and $V_{q}$, $Y_{q}$, and $Z_{q}$ be
    random variables for every $q \in \mathcal{Q}$.
    Let a random variable $Q$ take values in $\mathcal{Q}$ and be independent
    of $\{ ( V_{q} , Y_{q} , Z_{q} ) \}_{q \in \mathcal{Q}}$.
    If the support sets of $\{ Y_{q} \}_{q \in \mathcal{Q}}$ are disjoint, and
    $V_{q} \markovlink Y_{q} \markovlink Z_{q}$ for all $q \in \mathcal{Q}$,
    then $V_{Q} \markovlink Y_{Q} \markovlink Z_{Q}$.
\end{lemma}

Define functions $\{ \chi_{\lambda} \}_{\lambda \in \Lambda}$ from $\Lambda$ to
$\{ 0 , 1 \}$ by
\begin{equation}
    \chi_{\lambda} ( s )
    = \begin{cases}
        1 , & s \in \mathcal{S}_{\lambda} , \\
        0 , & s \not\in \mathcal{S}_{\lambda} .
    \end{cases} \label{e:allocation.indicator}
\end{equation}
Then $\Xi ( D )$ defined in Section~\ref{u:definitions.rdfunctions} equals the
set of random variables $\hat{X}$ satisfying
$S \markovlink X \markovlink \hat{X}$ and
\begin{equation}
    \mean{\chi_{\lambda} ( S ) d_{\lambda} ( X , \hat{X} )}
    \le \probability{S \in \mathcal{S}_{\lambda}} D ( \lambda )
    \label{e:allocation.gendis} % general distortion
\end{equation}
for all $\lambda \in \Lambda$.
To prove \eqref{e:results.properties.ratall}, we firstly prove that for every
$\hat{X} \in \Xi ( D )$, $\minfo{X ; \hat{X} | c ( X )}$ is no less than the
right-hand side of \eqref{e:results.properties.ratall}.
We then prove that
\begin{equation}
    R^{\mathrm{G}} ( D )
    \le \sum_{u \in \bar{\mathcal{U}}}
    \probability{c ( X ) = u}
    R_{u}^{*} ( \delta ( \cdot , u ) )
    + \epsilon \label{e:allocation.rates}
\end{equation}
for all $\delta \in \Delta ( D )$ and $\epsilon \in \oointerval{0}{\infty}$.

Let $\hat{X} \in \Xi ( D )$.
For $u \in \bar{\mathcal{U}}$, let $Q_{u}$ be the conditional distribution of
$( S , X , \hat{X} )$ given $c ( X ) = u$, and define a random tuple
$( T_{u} , Y_{u} , \hat{Y}_{u} )$ independent of $c ( X )$ and having the
distribution $Q_{u}$.
Then we have
\begin{equation}
    \minfo{X ; \hat{X} | c ( X )}
    = \sum_{u \in \bar{\mathcal{U}}}
    \probability{c ( X ) = u}
    \minfo{Y_{u} ; \hat{Y}_{u}}
    \label{e:allocation.cmi} % conditional mutual information
\end{equation}
and the total expectation formula
\begin{align}
    & \mean{\chi_{\lambda} ( S ) d_{\lambda} ( X , \hat{X} )} \notag \\
    & = \sum_{u \in \bar{\mathcal{U}}}
    \probability{c ( X ) = u}
    \mean{\chi_{\lambda} ( T_{u} ) d_{\lambda} ( Y_{u} , \hat{Y}_{u} )}
    \label{e:allocation.totex} % total expectation formula
\end{align}
for all $\lambda \in \Lambda$.
For $u \in \bar{\mathcal{U}}$, we have
$T_{u} \markovlink Y_{u} \markovlink \hat{Y}_{u}$ by
Lemma~\ref{lem:allocation.condition}.
There exists a function
$\delta : \Lambda \times \mathcal{U} \to \cointerval{0}{\infty}$ satisfying
\begin{equation}
    \mean{\chi_{\lambda} ( T_{u} ) d_{\lambda} ( Y_{u} , \hat{Y}_{u} )}
    = \probability{T_{u} \in \mathcal{S}_{\lambda}} \delta ( \lambda , u )
    \label{e:allocation.cde} % class distortion equation
\end{equation}
for all $\lambda \in \Lambda$ and $u \in \bar{\mathcal{U}}$: Note that
\eqref{e:allocation.cde} holds true even when
$\probability{T_{u} \in \mathcal{S}_{\lambda}} = 0$, because the left-hand side
of \eqref{e:allocation.cde} is zero in such a case.
Hence
\begin{equation}
    \minfo{Y_{u} ; \hat{Y}_{u}} \ge R_{u}^{*} ( \delta ( \cdot , u ) )
    \label{e:allocation.pcrate} % pessimistic class rate
\end{equation}
for all $u \in \bar{\mathcal{U}}$.
We can see $\delta \in \Delta ( D )$ from \eqref{e:allocation.gendis},
\eqref{e:allocation.totex}, and \eqref{e:allocation.cde}.
By \eqref{e:allocation.cmi}, \eqref{e:allocation.pcrate}, and
$\delta \in \Delta ( D )$, we conclude that $\minfo{X ; \hat{X} | c ( X )}$ is
no less than the right-hand side of \eqref{e:results.properties.ratall}.

Let $\delta \in \Delta ( D )$ and $\epsilon \in \oointerval{0}{\infty}$.
Since \eqref{e:allocation.rates} is apparent if
$R_{u}^{*} ( \delta ( \cdot , u ) ) = \infty$ for some
$u \in \bar{\mathcal{U}}$, we assume
$R_{u}^{*} ( \delta ( \cdot , u ) ) < \infty$ for all
$u \in \bar{\mathcal{U}}$.
Define a random pair $( T_{u} , Y_{u} ) \sim P_{u}$ for every
$u \in \bar{\mathcal{U}}$.
There exist random variables $\{ \hat{Y}_{u} \}_{u \in \bar{\mathcal{U}}}$
satisfying $T_{u} \markovlink Y_{u} \markovlink \hat{Y}_{u}$,
\begin{equation}
    \mean{\chi_{\lambda} ( T_{u} ) d_{\lambda} ( Y_{u} , \hat{Y}_{u} )}
    \le \probability{T_{u} \in \mathcal{S}_{\lambda}} \delta ( \lambda , u ) ,
    \label{e:allocation.cdi} % class distortion inequality
\end{equation}
and
\begin{equation}
    \minfo{Y_{u} ; \hat{Y}_{u}}
    < R_{u}^{*} ( \delta ( \cdot , u ) ) + \epsilon
    \label{e:allocation.ocrate} % optimistic class rate
\end{equation}
for all $\lambda \in \Lambda$ and $u \in \bar{\mathcal{U}}$.
Further define a random variable $U$ independent of
$\{ ( T_{u} , Y_{u} , \hat{Y}_{u} ) \}_{u \in \bar{\mathcal{U}}}$ and let $U$
and $c ( X )$ be identically distributed.
It is not difficult to verify that $( S , X , c ( X ) )$ and
$( T_{U} , Y_{U} , U )$ are identically distributed.
Consequently, we can define a random variable $\hat{X}$ such that
$( S , X , c ( X ) , \hat{X} )$ and $( T_{U} , Y_{U} , U , \hat{Y}_{U} )$ are
identically distributed.
Then \eqref{e:allocation.cmi} still holds, and \eqref{e:allocation.totex} still
holds for all $\lambda \in \Lambda$.
For $\lambda \in \Lambda$, \eqref{e:allocation.totex},
\eqref{e:allocation.cdi}, and \eqref{e:results.properties.disall} imply
\eqref{e:allocation.gendis}.
By Lemma~\ref{lem:allocation.mixture} we have
$T_{U} \markovlink Y_{U} \markovlink \hat{Y}_{U}$, which implies
$S \markovlink X \markovlink \hat{X}$.
Therefore,
\begin{equation}
    R^{\mathrm{G}} ( D ) \le \minfo{X ; \hat{X} | c ( X )} .
    \label{e:allocation.pessrate} % pessimistic rate
\end{equation}
The inequality \eqref{e:allocation.rates} follows from
\eqref{e:allocation.cmi}, \eqref{e:allocation.ocrate}, and
\eqref{e:allocation.pessrate}.

    \section{Proof of Theorem~\ref{thm:results.case.case}}
\label{u:cases}

We use Theorem~\ref{thm:results.properties.allocation} to prove
\eqref{e:results.case.conrate}.
Define $\bar{\mathcal{U}}$, $\{ P_{u} \}_{u \in \bar{\mathcal{U}}}$,
$\{ R_{u}^{*} \}_{u \in \bar{\mathcal{U}}}$, and $\Delta ( D )$ as in
Theorem~\ref{thm:results.properties.allocation}.
We can see $\Delta ( D )$ is the set of functions
$\delta : \Lambda \times \mathcal{U} \to \cointerval{0}{\infty}$ satisfying
\eqref{e:results.properties.disall} for all $\lambda \in \bar{\Lambda}$.
We have $\Lambda \cap \bar{\mathcal{U}} = \bar{\Lambda}$ because
$c ( X ) = \lambda$ is equivalent to $S \in \mathcal{S}_{\lambda}$ with
probability one for all $\lambda \in \Lambda$.
For $\delta : \Lambda \times \mathcal{U} \to \cointerval{0}{\infty}$ and
$\lambda \in \bar{\Lambda}$, \eqref{e:results.properties.disall} is equivalent
to $\delta ( \lambda , \lambda ) \le D ( \lambda )$.

Fix $u \in \bar{\mathcal{U}}$ and
$\tilde{D} : \Lambda \to \cointerval{0}{\infty}$.
Define a random pair $( T , Y ) \sim P_{u}$ and a set
$\tilde{\Lambda} = \{ \lambda \in \Lambda |
    \probability{T \in \mathcal{S}_{\lambda}} > 0
\}$.
Then
\begin{equation}
    R_{u}^{*} ( \tilde{D} )
    = \inf_{\hat{Y} \in \tilde{\Xi} ( \tilde{D} )} \minfo{Y ; \hat{Y}} ,
\end{equation}
where $\tilde{\Xi} ( \tilde{D} )$ is the set of random variables $\hat{Y}$
satisfying $T \markovlink Y \markovlink \hat{Y}$ and $
    \mean{d_{\lambda} ( Y , \hat{Y} ) | T \in \mathcal{S}_{\lambda}}
    \le \tilde{D} ( \lambda )
$ for all $\lambda \in \tilde{\Lambda}$.
Also note
\begin{align}
    \probability{T \in \mathcal{S}_{\lambda}}
    & = \probability{c ( X ) = \lambda | c ( X ) = u} \\
    & = \begin{cases}
        1 , & \lambda = u , \\
        0 , & \lambda \not= u
    \end{cases}
\end{align}
for $\lambda \in \Lambda$.
If $u \in \bar{\Lambda}$, then $Y \sim Q_{u}$,
$\probability{T \in \mathcal{S}_{u}} = 1$, $\tilde{\Lambda} = \{ u \}$, and
consequently $\tilde{\Xi} ( \tilde{D} )$ is the set of random variables
$\hat{Y}$ satisfying $T \markovlink Y \markovlink \hat{Y}$ and
$\mean{d_{u} ( Y , \hat{Y} )} \le \tilde{D} ( u )$.
If $u \not\in \bar{\Lambda}$, then $u \not\in \Lambda$,
$\tilde{\Lambda} = \emptyset$, and consequently $\tilde{\Xi} ( \tilde{D} )$ is
the set of random variables $\hat{Y}$ satisfying
$T \markovlink Y \markovlink \hat{Y}$.
Hence $R_{u}^{*} ( \tilde{D} ) = R^{( u )} ( \tilde{D} ( u ) )$ if
$u \in \bar{\Lambda}$, and $R_{u}^{*} ( \tilde{D} ) = 0$ otherwise.

Therefore,
\begin{equation}
    R^{\mathrm{G}} ( D )
    = \inf_{\delta \in \Delta ( D )}
    \sum_{u \in \bar{\Lambda}}
    \probability{c ( X ) = u}
    R^{( u )} ( \delta ( u , u ) ) ,
\end{equation}
and $\Delta ( D )$ is the set of functions
$\delta : \Lambda \times \mathcal{U} \to \cointerval{0}{\infty}$ satisfying
$\delta ( \lambda , \lambda ) \le D ( \lambda )$ for all
$\lambda \in \bar{\Lambda}$.
This proves \eqref{e:results.case.conrate}.

Now suppose $\bar{\Lambda}$ has only one element $\lambda$.
Fix $\epsilon \in \oointerval{0}{\infty}$.
Then there exists a random variable $\hat{X}$ satisfying
$S \markovlink X \markovlink \hat{X}$,
\begin{equation}
    \mean{d_{\lambda} ( X , \hat{X} ) | S \in \mathcal{S}_{\lambda}}
    \le D ( \lambda ) ,
    \label{e:case.dismean} % distortion mean
\end{equation}
and
\begin{equation}
    R^{\mathrm{G}} ( D ) + \epsilon > \minfo{X ; \hat{X} | c ( X )} .
    \label{e:case.conrate} % conditional rate
\end{equation}
Let $\hat{P}$ be the conditional distribution of $\hat{X}$ given
$c ( X ) = \lambda$.
Let a random variable $\hat{Y}$ have the distribution $\hat{P}$ and be
independent of $( S , X , \hat{X} )$.
Let a random variable $Z$ equals $\hat{X}$ when $c ( X ) = \lambda$, and equals
$\hat{Y}$ when $c ( X ) \not= \lambda$.
We will show $Z \in \Xi ( D )$, which implies an upper bound for $R^{*} ( D )$.
Subtracting $\minfo{S ; X}$ from both sides of the data processing inequality
$\minfo{S ; X , \hat{X} , \hat{Y}} \ge \minfo{S ; X , Z}$, we get
\begin{equation}
    \minfo{S ; \hat{X} | X} + \minfo{S ; \hat{Y} | X , \hat{X}}
    \ge \minfo{S ; Z | X} .
    \label{e:case.dpinequality} % data-processing inequality
\end{equation}
By the chain rule for mutual information,
\begin{equation}
    \minfo{S , X , \hat{X} ; \hat{Y}}
    = \minfo{X , \hat{X} ; \hat{Y}} + \minfo{S ; \hat{Y} | X , \hat{X}} .
    \label{e:case.micrule} % mutual information chain rule
\end{equation}
By \eqref{e:case.dpinequality}, \eqref{e:case.micrule}, and the fact
\begin{equation}
    \minfo{S ; \hat{X} | X}
    = \minfo{S , X , \hat{X} ; \hat{Y}}
    = \minfo{X , \hat{X} ; \hat{Y}}
    = 0 ,
\end{equation}
we have $S \markovlink X \markovlink Z$.
Also note \eqref{e:case.dismean} and
\begin{equation}
    \mean{d_{\lambda} ( X , Z ) | S \in \mathcal{S}_{\lambda}}
    = \mean{d_{\lambda} ( X , \hat{X} ) | S \in \mathcal{S}_{\lambda}} .
\end{equation}
By definition, we have $Z \in \Xi ( D )$, which implies
\begin{equation}
    R^{*} ( D ) \le \minfo{X ; Z} .
    \label{e:case.genrate} % general rate
\end{equation}
The data processing inequality
$\minfo{X ; \hat{X} , \hat{Y} , c ( X )} \ge \minfo{X ; Z , c ( X )}$, the
chain rule for mutual information, $\minfo{X ; \hat{Y}} = 0$, and
$\minfo{X ; \hat{X} , c ( X ) | \hat{Y}} = \minfo{X ; \hat{X} , c ( X )}$ lead
to
\begin{equation}
    \minfo{X ; \hat{X} , c ( X )} \ge \minfo{X ; Z , c ( X )} .
    \label{e:case.mutinf} % mutual information
\end{equation}
Subtracting $\minfo{X ; c ( X )}$ from \eqref{e:case.mutinf}, we get
\begin{equation}
    \minfo{X ; \hat{X} | c ( X )} \ge \minfo{X ; Z | c ( X )} .
    \label{e:case.cminf} % conditional mutual information
\end{equation}
For $u \in \bar{\mathcal{U}}$, $Z$ has the conditional distribution $\hat{P}$
given $c ( X ) = u$, no matter $u = \lambda$ or not.
Hence $c ( X )$ and $Z$ are independent.
Since $\minfo{c ( X ) ; Z} = 0$, $\minfo{c ( X ) ; Z | X} = 0$, and $
    \minfo{X ; Z | c ( X )} + \minfo{c ( X ) ; Z}
    = \minfo{X ; Z} + \minfo{c ( X ) ; Z | X}
$, we have
\begin{equation}
    \minfo{X ; Z | c ( X )} = \minfo{X ; Z} .
    \label{e:case.conremoving} % condition removing
\end{equation}
The inequality \eqref{e:case.conrate}, \eqref{e:case.genrate},
\eqref{e:case.cminf}, and \eqref{e:case.conremoving} lead to
$R^{\mathrm{G}} ( D ) + \epsilon > R^{*} ( D )$.
Since $\epsilon$ can be an arbitrary number in $\oointerval{0}{\infty}$, we
have proved $R^{\mathrm{G}} ( D ) \ge R^{*} ( D )$.

Therefore, $R^{\mathrm{G}} ( D ) = R^{*} ( D )$ if $| \bar{\Lambda} | = 1$.

    \section{Proofs of Lemmas}
    \label{s:lemmaproofs}
    \subsection{Lemma~\ref{lem:sketches.general.vlcc}}
\label{u:lemmaproofs.vlcc}

For $n \in \cointegers{1}{\infty}$, since $M_{n} \le 2^{\lceil n R \rceil}$,
there exist functions
$\alpha_{n} : \ocnumbers{M_{n}} \to \{ 0 , 1 \}^{\lceil n R \rceil}$ and
$\beta_{n} : \{ 0 , 1 \}^{\lceil n R \rceil} \to \ocnumbers{M_{n}}$ satisfying
$\beta_{n} ( \alpha_{n} ( m ) ) = m$ for all $m \in \ocnumbers{M_{n}}$.
Define $f : \mathcal{X}^{*} \to \{ 0 , 1 \}^{*}$ by
\begin{equation}
    f ( \boldsymbol{x} )
    = \begin{cases}
        \text{the empty string} , &
        | \boldsymbol{x} | = 0 , \\
        \gamma ( | \boldsymbol{x} | ) \alpha_{| \boldsymbol{x} |} (
            f_{| \boldsymbol{x} |} ( \boldsymbol{x} )
        ) , &
        | \boldsymbol{x} | > 0 ,
    \end{cases}
\end{equation}
where $\gamma ( | \boldsymbol{x} | )$ is the compound representation of
$| \boldsymbol{x} |$ (see Section~\ref{u:sketches.general}).
Since $\gamma$ is prefix-free, there exists a function
$\varphi : \{ 0 , 1 \}^{*} \to \hat{\mathcal{X}}^{*}$ that maps the empty
string to the empty string and satisfies $
    \varphi ( \gamma ( n ) \boldsymbol{b} )
    = \varphi_{n} ( \beta_{n} ( \boldsymbol{b} ) )
$ for all $n \in \cointegers{1}{\infty}$ and
$\boldsymbol{b} \in \{ 0 , 1 \}^{\lceil n R \rceil}$.
We can see $\varphi ( f ( \boldsymbol{x} ) ) = \varphi_{| \boldsymbol{x} |} (
    f_{| \boldsymbol{x} |} ( \boldsymbol{x} )
)$ for all non-empty strings $\boldsymbol{x} \in \mathcal{X}^{*}$.
Also note that $\varphi ( f ( \text{the empty string} ) )$ is the empty string.
Therefore, $( f , \varphi )$ is an
$( \mathcal{X} , \hat{\mathcal{X}} )$-variable-length code.
For $n \in \cointegers{1}{\infty}$ and $\boldsymbol{x} \in \mathcal{X}^{n}$, we
have $| \gamma ( n ) | \le 1 + 2 \log_{2} ( n )$,
$| \alpha_{n} ( f_{n} ( \boldsymbol{x} ) ) | < n R + 1$, and consequently
$| f ( \boldsymbol{x} ) | < 2 + 2 \log_{2} ( n ) + n R$.
This proves \eqref{e:sketches.general.rate}.

    \subsection{Lemma~\ref{lem:sketches.ctc.assemble}}
\label{u:lemmaproofs.assemble}

By basic information theory results, there exists a $\mathcal{U}$-lossless code
$( f_{\mathrm{L}} , \varphi_{\mathrm{L}} )$ such that
\begin{equation}
    \lim_{n \to \infty}
    \probability{| f_{\mathrm{L}} ( U_{\ocnumbers{n}} ) | \le n R_{\mathrm{L}}}
    = 1 \label{e:lemmaproofs.assemble.labelrate}
\end{equation}
for all $R_{L} > \entropy{c ( X )}$.
For $n \in \conumbers{\infty}$ and
$\{ x_{i} \}_{i = 1}^{n} \in \mathcal{X}^{n}$, let
\begin{equation}
    f_{2} ( x_{\ocnumbers{n}} )
    = \gamma ( | f_{\mathrm{L}} ( u_{\ocnumbers{n}} ) | + 1 )
    f_{\mathrm{L}} ( u_{\ocnumbers{n}} )
    f_{1} ( x_{\ocnumbers{n}} , u_{\ocnumbers{n}} ) ,
    \label{e:lemmaproofs.assemble.ctcencoder}
\end{equation}
where $\gamma$ denotes the compound representation introduced in
Section~\ref{u:sketches.general}, and $u_{i} = c ( x_{i} )$ for all
$i \in \ocnumbers{n}$.
Define $\varphi_{2} : \{ 0 , 1 \}^{*} \to \hat{\mathcal{X}}^{*}$ and
$\psi : \{ 0 , 1 \}^{*} \to \mathcal{U}^{*}$ such that
\begin{equation}
    \varphi_{2} ( \gamma ( l + 1 ) b_{\ocnumbers{m}} )
    = \varphi_{1} (
        b_{\ocintegers{l}{m}} ,
        \varphi_{\mathrm{L}} ( b_{\ocnumbers{l}} )
    )
\end{equation}
and $
    \psi ( \gamma ( l + 1 ) b_{\ocnumbers{m}} )
    = \varphi_{\mathrm{L}} ( b_{\ocnumbers{l}} )
$ hold for all $l \in \conumbers{\infty}$, $m \in \cointegers{l}{\infty}$, and
$\{ b_{i} \}_{i = 1}^{m} \in \{ 0 , 1 \}^{m}$.
This is possible because $\gamma$ is prefix-free.
If $n \in \conumbers{\infty}$, $\{ x_{i} \}_{i = 1}^{n} \in \mathcal{X}^{n}$,
and $u_{i} = c ( x_{i} )$ for all $i \in \ocnumbers{n}$, then we can verify
$\psi ( f_{2} ( x_{\ocnumbers{n}} ) ) = u_{\ocnumbers{n}}$ and
\eqref{e:sketches.ctc.idrep}.

Fix $R_{2} > R_{1} + \entropy{c ( X )}$.
Then there exists a value $R_{\mathrm{L}}$ satisfying
$R_{\mathrm{L}} > \entropy{c ( X )}$ and $R_{1} + R_{\mathrm{L}} < R_{2}$.
Equation~\eqref{e:sketches.ctc.ctcrate} follows from
\eqref{e:sketches.general.length_cr}, \eqref{e:sketches.ctc.parate},
\eqref{e:lemmaproofs.assemble.labelrate},
\eqref{e:lemmaproofs.assemble.ctcencoder}, and
$R_{1} + R_{\mathrm{L}} < R_{2}$.

    \section{Proof of Theorem~\ref{thm:results.theorems.general}}
    \label{s:genproof} % general proof
    \subsection{Direct Part}
\label{u:genproof.direct}

Let $R > R^{*} ( D )$.
Define functions
$\{ d_{\lambda , \delta} \}_{\lambda \in \Lambda , \delta \in \reals}$ by
\eqref{e:sketches.general.dismeas}.

By Proposition~1 in \cite{martinian2008} and
Corollary~\ref{cor:sketches.general.rdfunction}, there exist sequences
$\{ M_{n} \}_{n = 1}^{\infty}$, $\{ f_{n} \}_{n = 1}^{\infty}$, and
$\{ \varphi_{n} \}_{n = 1}^{\infty}$ such that
\begin{enumerate}
    \item For every $n \in \cointegers{1}{\infty}$,
    $M_{n} \le 2^{n R^{*} ( D )}$ is a positive integer, $f_{n}$ is a function
    from $\mathcal{X}^{n}$ to $\ocnumbers{M_{n}}$, and $\varphi_{n}$ is a
    function from $\ocnumbers{M_{n}}$ to $\hat{\mathcal{X}}^{n}$; and
    \item For every $\lambda \in \Lambda$ and
    $\epsilon \in \oointerval{0}{\infty}$, the probability that
    \begin{equation}
        \frac{1}{n}
        \sum_{i = 1}^{n}
        d_{\lambda , D ( \lambda )} ( X_{i} , \hat{X}_{n , i} , S_{i} )
        \le D ( \lambda ) + \epsilon
    \end{equation}
    tends to 1 as $n \to \infty$, where $\hat{X}_{n , i}$ is defined by $
        \{ \hat{X}_{n , i} \}_{i = 1}^{n}
        = \varphi_{n} ( f_{n} ( X_{\ocnumbers{n}} ) )
    $.
\end{enumerate}
By Lemma~\ref{lem:sketches.general.vlcc}, there exists an
$( \mathcal{X} , \hat{\mathcal{X}} )$-variable-length code $( f , \varphi )$
satisfying \eqref{e:results.theorems.rate} and $
    \varphi ( f ( \boldsymbol{x} ) )
    = \varphi_{| \boldsymbol{x} |} (
        f_{| \boldsymbol{x} |} ( \boldsymbol{x} )
    )
$ for all non-empty strings $\boldsymbol{x} \in \mathcal{X}^{*}$.
Hence there exists $n_{1} \in \cointegers{1}{\infty}$ such that
$| f ( \boldsymbol{x} ) | \le n R$ for all integers $n > n_{1}$ and
$\boldsymbol{x} \in \mathcal{X}^{n}$.
We then have
\begin{equation}
    \lim_{n \to \infty} \probability{| f ( X_{\ocnumbers{n}} ) | \le n R} = 1 .
    \label{e:genproof.direct.slprob} % small-length probability
\end{equation}

We now analyze the distortion of $( f , \varphi )$.
Fix $\lambda \in \bar{\Lambda}$ and $\epsilon \in \oointerval{0}{\infty}$.
Since $\probability{S \in \mathcal{S}_{\lambda}} > 0$, the probability that
\begin{equation}
    \frac{1}{n}
    \sum_{i = 1}^{n}
    d_{\lambda , D ( \lambda )} ( X_{i} , \hat{X}_{n , i} , S_{i} )
    \le D ( \lambda )
    + \frac{\epsilon}{2} \probability{S \in \mathcal{S}_{\lambda}}
    \label{e:genproof.direct.exresult} % existing result
\end{equation}
tends to 1 as $n \to \infty$.
Define a function $\chi_{\lambda} : \mathcal{S} \to \{ 0 , 1 \}$ by
\eqref{e:allocation.indicator}.
Then
\begin{align}
    d_{\lambda , D ( \lambda ) + \epsilon} ( X_{i} , \hat{X}_{n , i} , S_{i} )
    ={} & d_{\lambda , D ( \lambda )} ( X_{i} , \hat{X}_{n , i} , S_{i} )
    \notag \\
    & + \epsilon
    - \epsilon \chi_{\lambda} ( S_{i} )
    \label{e:genproof.direct.disdiff} % distortion difference
\end{align}
with probability one for all integers $n$ and $i$ satisfying $0 < i \le n$.
By the law of large numbers, the probability that
\begin{equation}
    \frac{1}{n} \sum_{i = 1}^{n} \chi_{\lambda} ( S_{i} )
    \ge \frac{1}{2} \probability{S \in \mathcal{S}_{\lambda}}
    \label{e:genproof.direct.wlln} % the weak law of large numbers
\end{equation}
tends to 1 as $n \to \infty$.
If \eqref{e:genproof.direct.disdiff} holds for all $i \in \ocnumbers{n}$ with
$n \in \cointegers{1}{\infty}$, \eqref{e:genproof.direct.exresult} holds, and
\eqref{e:genproof.direct.wlln} holds, then we have
\begin{equation}
    \frac{1}{n}
    \sum_{i = 1}^{n}
    d_{\lambda , D ( \lambda ) + \epsilon}
    ( X_{i} , \hat{X}_{n , i} , S_{i} )
    \le D ( \lambda ) + \epsilon ,
\end{equation}
which implies that $S_{\ocnumbers{n}}$, $X_{\ocnumbers{n}}$, and
$\varphi ( f ( X_{\ocnumbers{n}} ) )$ meet the fidelity criterion
$( d_{\lambda} , \mathcal{S}_{\lambda} , D ( \lambda ) + \epsilon )$.
Therefore, the probability with which $S_{\ocnumbers{n}}$, $X_{\ocnumbers{n}}$,
and $\varphi ( f ( X_{\ocnumbers{n}} ) )$ meet the fidelity criterion
$( d_{\lambda} , \mathcal{S}_{\lambda} , D ( \lambda ) + \epsilon )$ tends to 1
as $n \to \infty$.

For $\lambda \in \Lambda \setminus \bar{\Lambda}$,
$\epsilon \in \oointerval{0}{\infty}$, and $n \in \cointegers{1}{\infty}$,
because $S_{i} \not\in \mathcal{S}_{\lambda}$ with probability one for all
$i \in \ocnumbers{n}$, the probability with which $S_{\ocnumbers{n}}$,
$X_{\ocnumbers{n}}$, and $\varphi ( f ( X_{\ocnumbers{n}} ) )$ meet the
fidelity criterion
$( d_{\lambda} , \mathcal{S}_{\lambda} , D ( \lambda ) + \epsilon )$ is 1.
In conclusion, this probability tends to 1 as $n \to \infty$ for all
$\lambda \in \Lambda$ and $\epsilon \in \oointerval{0}{\infty}$.
This fact and \eqref{e:genproof.direct.slprob} lead to the conclusion that
$( f , \varphi )$ achieves $( R , D )$ for $( S , X )$ and
$\{ ( d_{\lambda} , \mathcal{S}_{\lambda} ) \}_{\lambda \in \Lambda}$.

    \subsection{Converse Part}
\label{u:genproof.converse}

Suppose an $( \mathcal{X} , \hat{\mathcal{X}} )$-variable-length code
$( f , \varphi )$ achieves $( R , D )$ for $( S , X )$ and
$\{ ( d_{\lambda} , \mathcal{S}_{\lambda} ) \}_{\lambda \in \Lambda}$.

We firstly fix an arbitrary $n \in \cointegers{1}{\infty}$ and define a block code.
Let $M_{n}$ be the number of the bit strings in $f ( \mathcal{X}^{n} )$ whose
length does not exceed $n R$.
There exists functions $f_{n} : \mathcal{X}^{n} \to \ocnumbers{M_{n}}$ and
$\varphi_{n} : \ocnumbers{M_{n}} \to \hat{\mathcal{X}}^{n}$ such that for every
$\boldsymbol{x} \in \mathcal{X}^{n}$, if $| f ( \boldsymbol{x} ) | \le n R$,
then
$\varphi_{n} ( f_{n} ( \boldsymbol{x} ) ) = \varphi ( f ( \boldsymbol{x} ) )$.
Since
\begin{equation}
    M_{n}
    \le \sum_{l = 0}^{\lfloor n R \rfloor} 2^{l}
    = 2^{\lfloor n R \rfloor + 1} - 1
    \le 2^{n R + 1} ,
\end{equation}
the rate $\log_{2} ( M_{n} ) / n$ of the block code $( f_{n} , \varphi_{n} )$
does not exceed $R + 1 / n$.

Now we analyze the distortion of the block codes.
Fix $\lambda \in \Lambda$ and $\epsilon \in \oointerval{0}{\infty}$.
For $n \in \cointegers{1}{\infty}$, define the following sets:
\begin{itemize}
    \item $F_{n}$ --- the set of $
        ( \boldsymbol{s} , \boldsymbol{x} )
        \in \mathcal{S}^{n} \times \mathcal{X}^{n}$
    such that $\boldsymbol{s}$, $\boldsymbol{x}$, and
    $\varphi ( f ( \boldsymbol{x} ) )$ meet the fidelity criterion
    $( d_{\lambda} , \mathcal{S}_{\lambda} , D ( \lambda ) + \epsilon )$; and
    \item $G_{n}$ --- the set of $
        ( \{ s_{i} \}_{i = 1}^{n} , \{ x_{i} \}_{i = 1}^{n} )
        \in \mathcal{S}^{n} \times \mathcal{X}^{n}$
    such that $
        \{ \hat{x}_{i} \}_{i = 1}^{n}
        = \varphi_{n} ( f_{n} ( x_{\ocnumbers{n}} ) )
    $ satisfies
    \begin{equation}
        \frac{1}{n}
        \sum_{i = 1}^{n}
        d_{\lambda , D ( \lambda )} ( x_{i} , \hat{x}_{i} , s_{i} )
        \le D ( \lambda ) + \epsilon .
    \end{equation}
\end{itemize}
For $n \in \cointegers{1}{\infty}$ and
$( \{ s_{i} \}_{i = 1}^{n} , \{ x_{i} \}_{i = 1}^{n} )$, if
$( s_{\ocnumbers{n}} , x_{\ocnumbers{n}} ) \in F_{n}$,
$| f ( x_{\ocnumbers{n}} ) | \le n R$, and
$\{ \hat{x}_{i} \}_{i = 1}^{n} = \varphi_{n} ( f_{n} ( x_{\ocnumbers{n}} ) )$,
then we have
$\{ \hat{x}_{i} \}_{i = 1}^{n} = \varphi ( f ( x_{\ocnumbers{n}} ) )$,
\begin{equation}
    \frac{1}{n}
    \sum_{i = 1}^{n}
    d_{\lambda , D ( \lambda ) + \epsilon} ( x_{i} , \hat{x}_{i} , s_{i} )
    \le D ( \lambda ) + \epsilon ,
\end{equation}
and consequently $( s_{\ocnumbers{n}} , x_{\ocnumbers{n}} ) \in G_{n}$ by the
discussions in Section~\ref{u:sketches.general} and the fact that
\begin{equation}
    d_{\lambda , D ( \lambda )} ( x , \hat{x} , s )
    \le d_{\lambda , D ( \lambda ) + \epsilon} ( x , \hat{x} , s )
\end{equation}
for all $x \in \mathcal{X}$, $\hat{x} \in \hat{\mathcal{X}}$, and
$s \in \mathcal{S}$.
Hence for $n \in \cointegers{1}{\infty}$, we have
$( S_{\ocnumbers{n}} , X_{\ocnumbers{n}} ) \in G_{n}$ whenever
$( S_{\ocnumbers{n}} , X_{\ocnumbers{n}} ) \in F_{n}$ and
$| f ( X_{\ocnumbers{n}} ) | \le n R$.
Also note that
$\probability{( S_{\ocnumbers{n}} , X_{\ocnumbers{n}} ) \in F_{n}}$ and
$\probability{| f ( X_{\ocnumbers{n}} ) | \le n R}$ tend to 1 as
$n \to \infty$.
Therefore
\begin{equation}
    \lim_{n \to \infty}
    \probability{( S_{\ocnumbers{n}} , X_{\ocnumbers{n}} ) \in G_{n}}
    = 1 .
\end{equation}

The inequality $R \ge R^{*} ( D )$ follows from Proposition~1 in
\cite{martinian2008} and Corollary~\ref{cor:sketches.general.rdfunction}.

    \section{%
    Proof of Theorems~\ref{thm:results.theorems.ctc} and
    \ref{thm:results.theorems.condition}%
}
\label{s:ctcproofs}

In Section~\ref{u:sketches.ctc} we have seen that the converse part of
Theorem~\ref{thm:results.theorems.ctc} implies that of
Theorem~\ref{thm:results.theorems.condition}.
Also, the direct part of Theorem~\ref{thm:results.theorems.condition} implies
that of Theorem~\ref{thm:results.theorems.ctc}.
Therefore, we only need to prove the converse part of
Theorem~\ref{thm:results.theorems.ctc} and the direct part of
Theorem~\ref{thm:results.theorems.condition}.

\subsection{Converse Part of Theorem~\ref{thm:results.theorems.ctc}}
\label{u:ctcproofs.ctcconverse}

Suppose an $( \mathcal{X} , \hat{\mathcal{X}} )$-variable-length code
$( f , \varphi )$ achieves $( R , D )$ for $( S , X )$ and
$\{ ( d_{\lambda} , \mathcal{S}_{\lambda} ) \}_{\lambda \in \Lambda}$, $\psi$
is a function from $\{ 0 , 1 \}^{*}$ to $\mathcal{U}^{*}$, and
$\psi ( f ( X_{\ocnumbers{n}} ) ) = U_{\ocnumbers{n}}$ with probability one for
all $n \in \conumbers{\infty}$.

Define $\tilde{\lambda}$, $\tilde{\Lambda}$,
$\{ \tilde{d}_{\lambda} \}_{\lambda \in \tilde{\Lambda}}$,
$\{ \tilde{\mathcal{S}}_{\lambda} \}_{\lambda \in \tilde{\Lambda}}$,
$\tilde{R}$, and $\tilde{D}$ as in
Corollary~\ref{cor:results.properties.admeasure}.
For $\boldsymbol{b} \in \{ 0 , 1 \}^{*}$, let $
    \tilde{\varphi} ( \boldsymbol{b} )
    = \{ ( \hat{x}_{i} , \hat{u}_{i} ) \}_{i = 1}^{n}
$, where $\{ \hat{x}_{i} \}_{i = 1}^{n} = \varphi ( \boldsymbol{b} )$ and
$\{ \hat{u}_{i} \}_{i = 1}^{n} = \psi ( \boldsymbol{b} )$.

Fix $\epsilon \in \oointerval{0}{\infty}$.
For $\lambda \in \Lambda$ and $n \in \conumbers{\infty}$, let $p_{\lambda , n}$
denote the probability with which $S_{\ocnumbers{n}}$, $X_{\ocnumbers{n}}$, and
$\varphi ( f ( X_{\ocnumbers{n}} ) )$ meet the fidelity criterion
$( d_{\lambda} , \mathcal{S}_{\lambda} , D ( \lambda ) + \epsilon )$.
For $\lambda \in \tilde{\Lambda}$ and $n \in \conumbers{\infty}$, let
$\tilde{p}_{\lambda , n}$ denote the probability with which
$S_{\ocnumbers{n}}$, $X_{\ocnumbers{n}}$, and
$\tilde{\varphi} ( f ( X_{\ocnumbers{n}} ) )$ meet the fidelity criterion $(
    \tilde{d}_{\lambda} ,
    \tilde{\mathcal{S}}_{\lambda} ,
    \tilde{D} ( \lambda ) + \epsilon
)$.
Then for $\lambda \in \Lambda$,
$\tilde{p}_{\lambda , n} = p_{\lambda , n} \to 1$ as $n \to \infty$.
For $n \in \conumbers{\infty}$, because
$\psi ( f ( X_{\ocnumbers{n}} ) ) = U_{\ocnumbers{n}}$ with probability one,
$\tilde{p}_{\tilde{\lambda} , n} = 1$.
In short, $\tilde{p}_{\lambda , n} \to 1$ as $n \to \infty$ for all
$\lambda \in \tilde{\Lambda}$.

Also note that $\probability{| f ( X_{\ocnumbers{n}} ) | \le n R} \to 1$ as
$n \to \infty$.
Therefore, the
$( \mathcal{X} , \hat{\mathcal{X}} \times \mathcal{U} )$-variable-length code
$( f , \tilde{\varphi} )$ achieves $( R , \tilde{D} )$ for $( S , X )$ and $\{
    ( \tilde{d}_{\lambda} , \tilde{\mathcal{S}}_{\lambda} )
\}_{\lambda \in \tilde{\Lambda}}$.
By the converse part of Theorem~\ref{thm:results.theorems.general}, we have
$R \ge \tilde{R} ( \tilde{D} )$, i.e. $R \ge R^{\mathrm{C}} ( D )$.

\subsection{Direct Part of Theorem~\ref{thm:results.theorems.condition}}
\label{u:ctcproofs.condirect} % parallel direct

Suppose $R > R^{\mathrm{G}} ( D )$.

Define $\bar{\mathcal{U}}$, $\{ P_{u} \}_{u \in \bar{\mathcal{U}}}$, and
$\{ R_{u}^{*} \}_{u \in \bar{\mathcal{U}}}$ as in
Theorem~\ref{thm:results.properties.allocation}.
By Theorem~\ref{thm:results.properties.allocation}, there exists a function
$\delta : \Lambda \times \mathcal{U} \to \cointerval{0}{\infty}$ satisfying $
    \sum_{u \in \bar{\mathcal{U}}}
    \probability{c ( X ) = u}
    R_{u}^{*} ( \delta ( \cdot , u ) )
    < R
$ and \eqref{e:results.properties.disall} for all $\lambda \in \Lambda$.
Then there exist $\{ q_{u} \}_{u \in \mathcal{U}}$ and
$\{ R_{u} \}_{u \in \mathcal{U}}$ satisfying
\begin{equation}
    \sum_{u \in \bar{\mathcal{U}}} q_{u} R_{u} < R ,
    \label{e:ctcproofs.condirect.rates}
\end{equation}
$\probability{c ( X ) = u} < q_{u}$, and
$R_{u}^{*} ( \delta ( \cdot , u ) ) < R_{u}$ for all $u \in \bar{\mathcal{U}}$.
For $u \in \bar{\mathcal{U}}$, by the direct part of
Theorem~\ref{thm:results.theorems.general}, there exists an
$( \mathcal{X} , \hat{\mathcal{X}} )$-variable-length code
$( f_{u} , \varphi_{u} )$ achieving $( R_{u} , \delta ( \cdot , u ) )$ for a
$P_{u}$-distributed random pair (see
Theorem~\ref{thm:results.properties.allocation}) and
$\{ ( d_{\lambda} , \mathcal{S}_{\lambda} ) \}_{\lambda \in \Lambda}$.
For $u \in \mathcal{U} \setminus \bar{\mathcal{U}}$, define
$f_{u} : \mathcal{X}^{*} \to \{ 0 , 1 \}^{*}$ by
$f_{u} ( \boldsymbol{x} ) = \gamma ( | \boldsymbol{x} | + 1 )$, and let
$\varphi_{u} : \{ 0 , 1 \}^{*} \to \hat{\mathcal{X}}^{*}$ satisfy
$\varphi_{u} ( \gamma ( k + 1 ) ) \in \hat{\mathcal{X}}^{k}$ for all
$k \in \conumbers{\infty}$, where $\gamma$ is the compound representation
introduced in Section~\ref{u:sketches.general}.
Then
$\{ ( f_{u} , \varphi_{u} ) \}_{u \in \mathcal{U}}$ are
$( \mathcal{X} , \hat{\mathcal{X}} )$-variable-length codes.
These variable-length codes will be combined to form the desired label-based
code.
Let $\mu ( i )$ denote the $i$-th element of the set $\mathcal{U}$: That is, $
    \mathcal{U}
    = \{ \mu ( 1 ) , \mu ( 2 ) , \cdots , \mu ( | \mathcal{U} | ) \}
$.
For $n \in \conumbers{\infty}$, $\{ x_{i} \}_{i = 1}^{n} \in \mathcal{X}^{n}$,
and $\{ u_{i} \}_{i = 1}^{n} \in \mathcal{U}^{n}$, let $
    f ( x_{\ocnumbers{n}} , u_{\ocnumbers{n}} )
    = \boldsymbol{b}_{\mu ( 1 )}
    \boldsymbol{b}_{\mu ( 2 )}
    \cdots
    \boldsymbol{b}_{\mu ( | \mathcal{U} | )}
$, where $J ( u ) = \{ i \in \ocnumbers{n} | u_{i} = u \}$ and
\begin{equation}
    \boldsymbol{b}_{u}
    = \begin{cases}
        \gamma ( | f_{u} ( x_{J ( u )} ) | + 1 ) f_{u} ( x_{J ( u )} ) , &
        u \in \bar{\mathcal{U}} , \\
        f_{u} ( x_{J ( u )} ) , &
        u \not\in \bar{\mathcal{U}}
    \end{cases}
\end{equation}
for all $u \in \mathcal{U}$.
Since $\gamma$ is prefix-free, the bit strings
$\{ f_{u} ( x_{J ( u )} ) \}_{u \in \mathcal{U}}$ can be recovered from
$f ( x_{\ocnumbers{n}} , u_{\ocnumbers{n}} )$.
Hence there exists a function
$\varphi : \{ 0 , 1 \}^{*} \times \mathcal{U}^{*} \to \hat{\mathcal{X}}^{*}$
such that
\begin{enumerate}
    \item $( f , \varphi )$ is an
    $( \mathcal{X} , \hat{\mathcal{X}} , \mathcal{U} )$-label-based code; and
    \item If $n \in \conumbers{\infty}$ and \eqref{e:results.theorems.sicoding}
    hold, then $\hat{x}_{J ( u )} = \varphi_{u} ( f_{u} ( x_{J ( u )} ) )$ for
    all $u \in \mathcal{U}$, where
    $J ( u ) = \{ i \in \ocnumbers{n} | u_{i} = u \}$.
\end{enumerate}
Define
\begin{equation}
    \{ \hat{X}_{i}^{( n )} \}_{i = 1}^{n}
    = \varphi (
        f ( X_{\ocnumbers{n}} , U_{\ocnumbers{n}} ) ,
        U_{\ocnumbers{n}}
    ) \label{e:ctcproofs.condirect.reproduction}
\end{equation}
and $J ( n , u ) = \{ i \in \ocnumbers{n} | U_{i} = u \}$ for
$n \in \cointegers{1}{\infty}$ and $u \in \mathcal{U}$.
Then
\begin{align}
    & | f ( X_{\ocnumbers{n}} , U_{\ocnumbers{n}} ) |
    = \sum_{u \in \mathcal{U} \setminus \bar{\mathcal{U}}}
    | \gamma ( | J ( n , u ) | + 1 ) | \notag \\
    & + \sum_{u \in \bar{\mathcal{U}}}
    | \gamma ( | f_{u} ( X_{J ( n , u )} ) | + 1 ) |
    + \sum_{u \in \bar{\mathcal{U}}} | f_{u} ( X_{J ( n , u )} ) |
    \label{e:ctcproofs.condirect.lengths}
\end{align}
with probability one for all $n \in \cointegers{1}{\infty}$, and
\begin{equation}
    \hat{X}_{J ( n , u )}^{( n )} = \varphi_{u} ( f_{u} ( X_{J ( n , u )} ) )
    \label{e:ctcproofs.condirect.classcoding}
\end{equation}
with probability one for all $n \in \cointegers{1}{\infty}$ and
$u \in \mathcal{U}$.

We now analyze the performance of $( f , \varphi )$.
Fix $u \in \bar{\mathcal{U}}$ and $\epsilon \in \oointerval{0}{\infty}$.
Define random pairs $\{ ( T_{i} , Y_{i} ) \}_{i = 1}^{\infty}$ as in
Lemma~\ref{lem:sketches.ctc.distribution}.
Then
\begin{equation}
    \lim_{k \to \infty} \probability{
        ( T_{\ocnumbers{k}} , Y_{\ocnumbers{k}} ) \in F_{u , \epsilon}
    }
    = 1 ,
\end{equation}
where $F_{u , \epsilon}$ is the set of $
    ( \boldsymbol{s} , \boldsymbol{x} )
    \in ( \mathcal{S} \times \mathcal{X} )^{*}
$ such that $| f_{u} ( \boldsymbol{x} ) | \le | \boldsymbol{x} | R_{u}$ and
that $\boldsymbol{s}$, $\boldsymbol{x}$, and
$\varphi_{u} ( f_{u} ( \boldsymbol{x} ) )$ meet the fidelity criterion
$( d_{\lambda} , \mathcal{S}_{\lambda} , \delta ( \lambda , u ) + \epsilon )$
for all $\lambda \in \Lambda$.
By Lemma~\ref{lem:sketches.ctc.asymlength} we have
\begin{equation}
    \lim_{n \to \infty}
    \probability{( T_{J ( n , u )} , Y_{J ( n , u )} ) \in F_{u , \epsilon}}
    = 1 .
\end{equation}
By Lemma~\ref{lem:sketches.ctc.distribution},
$( S_{J ( n , u )} , X_{J ( n , u )} )$ and
$( T_{J ( n , u )} , Y_{J ( n , u )} )$ are identically distributed for all
$n \in \cointegers{1}{\infty}$.
Hence
\begin{equation}
    \lim_{n \to \infty}
    \probability{( S_{J ( n , u )} , X_{J ( n , u )} ) \in F_{u , \epsilon}}
    = 1 .
\end{equation}
By the law of large numbers, the probability of $| J ( n , u ) | \le n q_{u}$
also tends to 1 as $n \to \infty$.
Hence
\begin{equation}
    \lim_{n \to \infty}
    \probability{| f_{u} ( X_{J ( n , u )} ) | \le n q_{u} R_{u}}
    = 1 . \label{e:ctcproofs.condirect.classlength}
\end{equation}

Fix $\lambda \in \bar{\Lambda}$ and $\epsilon \in \oointerval{0}{\infty}$.
Define $J_{n} = \{ i \in \ocnumbers{n} | S_{i} \in \mathcal{S}_{\lambda} \}$
for $n \in \cointegers{1}{\infty}$.
By \eqref{e:results.properties.disall} and the law of large numbers, the
probability that
\begin{equation}
    \sum_{u \in \mathcal{U}}
    | J ( n , u ) \cap J_{n} |
    \left( \delta ( \lambda , u ) + \frac{\epsilon}{2} \right)
    \le | J_{n} | ( D ( \lambda ) + \epsilon )
\end{equation}
tends to 1 as $n \to \infty$.
Also note that, for every $u \in \mathcal{U}$, the probability with which
$S_{J ( n , u )}$, $X_{J ( n , u )}$, and
$\varphi_{u} ( f_{u} ( X_{J ( n , u )} ) )$ meet the fidelity criterion $(
    d_{\lambda} ,
    \mathcal{S}_{\lambda} ,
    \delta ( \lambda , u ) + \epsilon / 2
)$ tends to 1 as $n \to \infty$.
By \eqref{e:ctcproofs.condirect.reproduction},
\eqref{e:ctcproofs.condirect.classcoding}, and
Lemma~\ref{lem:sketches.ctc.fidelity}, the probability with which
$S_{\ocnumbers{n}}$, $X_{\ocnumbers{n}}$, and
$\varphi ( f ( X_{\ocnumbers{n}} , U_{\ocnumbers{n}} ) , U_{\ocnumbers{n}} )$
meet the fidelity criterion
$( d_{\lambda} , \mathcal{S}_{\lambda} , D ( \lambda ) + \epsilon )$ tends to 1
as $n \to \infty$.

Furthermore, $
    \lim_{n \to \infty}
    \probability{| f ( X_{\ocnumbers{n}} , U_{\ocnumbers{n}} ) | \le n R}
    = 1
$ follows from \eqref{e:sketches.general.length_cr},
\eqref{e:ctcproofs.condirect.rates}, \eqref{e:ctcproofs.condirect.lengths}, and
\eqref{e:ctcproofs.condirect.classlength}.
Therefore, $( f , \varphi )$ achieves $( R , D )$ for $( S , X )$ and
$\{ ( d_{\lambda} , \mathcal{S}_{\lambda} ) \}_{\lambda \in \Lambda}$ given
$c ( X )$.

    \bibliography{IEEEabrv, bibliography}

\begin{thebibliography}{10}
\providecommand{\url}[1]{#1}
\csname url@samestyle\endcsname
\providecommand{\newblock}{\relax}
\providecommand{\bibinfo}[2]{#2}
\providecommand{\BIBentrySTDinterwordspacing}{\spaceskip=0pt\relax}
\providecommand{\BIBentryALTinterwordstretchfactor}{4}
\providecommand{\BIBentryALTinterwordspacing}{\spaceskip=\fontdimen2\font plus
\BIBentryALTinterwordstretchfactor\fontdimen3\font minus
  \fontdimen4\font\relax}
\providecommand{\BIBforeignlanguage}[2]{{%
\expandafter\ifx\csname l@#1\endcsname\relax
\typeout{** WARNING: IEEEtran.bst: No hyphenation pattern has been}%
\typeout{** loaded for the language `#1'. Using the pattern for}%
\typeout{** the default language instead.}%
\else
\language=\csname l@#1\endcsname
\fi
#2}}
\providecommand{\BIBdecl}{\relax}
\BIBdecl

\bibitem{shannon1948}
C.~Shannon, ``A mathematical theory of communication,'' \emph{Bell Syst. Tech.
  J.}, vol.~27, pp. 379--423 and 623--656, Jul. 1948.

\bibitem{shannon1959}
------, ``Coding theorems for a discrete source with a fidelity criterion,''
  \emph{IRE Conv. Rec.}, 1959.

\bibitem{berger1971theory}
T.~Berger, \emph{Rate Distortion Theory}.\hskip 1em plus 0.5em minus
  0.4em\relax Englewood Cliffs, NJ, USA: Prentice-Hall, 1971.

\bibitem{cover2006}
T.~Cover and J.~Thomas, \emph{Elements of Information Theory}, 2nd~ed.\hskip
  1em plus 0.5em minus 0.4em\relax Hoboken, New Jersey: John Wiley \& Sons,
  2006.

\bibitem{csiszar2011}
I.~Csiszár and J.~Körner, \emph{Information Theory: Coding Theorems for
  Discrete Memoryless Systems}, 2nd~ed.\hskip 1em plus 0.5em minus 0.4em\relax
  Cambridge University Press, 2011.

\bibitem{szeliski2011}
R.~Szeliski, \emph{Computer Vision}.\hskip 1em plus 0.5em minus 0.4em\relax
  Springer, 2011.

\bibitem{rabiner2007}
L.~Rabiner and R.~Schafer, ``Introduction to digital speech processing,''
  \emph{Found. Trends Signal Process.}, vol.~1, no. 1-2, pp. 1--194, Dec. 2007.

\bibitem{christopoulos2000}
C.~Christopoulos, A.~Skodras, and T.~Ebrahimi, ``The {JPEG2000} still image
  coding system: An overview,'' \emph{{IEEE} Trans. Consum. Electron.},
  vol.~46, no.~4, pp. 1103--1127, Nov. 2000.

\bibitem{gray1972}
R.~Gray, ``Conditional rate-distortion theory,'' Information System Laboratory,
  Stanford University, Stanford, California, Tech. Rep. 6502-2, Oct. 1972.

\bibitem{gray1973}
------, ``A new class of lower bounds to information rates of stationary
  sources via conditional rate-distortion functions,'' \emph{{IEEE} Trans. Inf.
  Theory}, vol.~19, no.~4, pp. 480--489, Jul. 1973.

\bibitem{wyner1976}
A.~Wyner and J.~Ziv, ``The rate-distortion function for source coding with side
  information at the decoder,'' \emph{{IEEE} Trans. Inf. Theory}, vol.~22,
  no.~1, pp. 1--10, Jan. 1976.

\bibitem{berger1971game}
T.~Berger, ``The source coding game,'' \emph{{IEEE} Trans. Inf. Theory},
  vol.~17, no.~1, pp. 71--76, Jan. 1971.

\bibitem{linder2000}
T.~Linder, R.~Zamir, and K.~Zeger, ``On source coding with
  side-information-dependent distortion measures,'' \emph{{IEEE} Trans. Inf.
  Theory}, vol.~46, no.~7, pp. 2697--2704, Nov. 2000.

\bibitem{martinian2008}
E.~Martinian, G.~Wornell, and R.~Zamir, ``Source coding with distortion side
  information,'' \emph{{IEEE} Trans. Inf. Theory}, vol.~54, no.~10, pp.
  4638--4665, Oct. 2008.

\bibitem{dobrushin1962}
R.~Dobrushin and B.~Tsybakov, ``Information transmission with additional
  noise,'' \emph{IRE Trans. Inf. Theory}, vol.~8, no.~5, pp. 293--304, Sep.
  1962.

\bibitem{wolf1970}
J.~Wolf and J.~Ziv, ``Transmission of noisy information to a noisy receiver
  with minimum distortion,'' \emph{{IEEE} Trans. Inf. Theory}, vol.~16, no.~4,
  pp. 406--411, Jul. 1970.

\bibitem{witsenhausen1980}
H.~Witsenhausen, ``Indirect rate distortion problems,'' \emph{{IEEE} Trans.
  Inf. Theory}, vol.~26, no.~5, pp. 518--521, Sep. 1980.

\bibitem{fontana1980}
R.~Fontana, ``Universal codes for a class of composite sources (corresp.),''
  \emph{{IEEE} Trans. Inf. Theory}, vol.~26, no.~4, pp. 480--482, Jul. 1980.

\bibitem{rosenthal1988}
H.~Rosenthal and J.~Binia, ``On the epsilon entropy of mixed random
  variables,'' \emph{{IEEE} Trans. Inf. Theory}, vol.~34, no.~5, pp.
  1110--1114, Sep. 1988.

\bibitem{gyorgy1999}
A.~Gyorgy, T.~Linder, and K.~Zeger, ``On the rate-distortion function of random
  vectors and stationary sources with mixed distributions,'' \emph{{IEEE}
  Trans. Inf. Theory}, vol.~45, no.~6, pp. 2110--2115, Sep. 1999.

\bibitem{weidmann2012}
C.~Weidmann and M.~Vetterli, ``Rate distortion behavior of sparse sources,''
  \emph{{IEEE} Trans. Inf. Theory}, vol.~58, no.~8, pp. 4969--4992, Aug. 2012.

\bibitem{gray1974}
R.~Gray and L.~Davisson, ``Source coding theorems without the ergodic
  assumption,'' \emph{{IEEE} Trans. Inf. Theory}, vol.~20, no.~4, pp. 502--516,
  Jul. 1974.

\bibitem{kalveram1989}
H.~Kalveram and P.~Meissner, ``{Itakura-Saito} clustering and rate distortion
  functions for a composite source model of speech,'' \emph{Signal Processing},
  vol.~18, no.~2, pp. 195--216, Oct. 1989.

\bibitem{gibson2014}
J.~Gibson and J.~Hu, ``Rate distortion bounds for voice and video,''
  \emph{Found. Trends Commun. Inf. Theory}, vol.~10, no.~4, pp. 379--514, Feb.
  2014.

\bibitem{gibson2017}
J.~Gibson, ``Rate distortion functions and rate distortion function lower
  bounds for real-world sources,'' \emph{Entropy}, vol.~19, no.~11, pp. 1--22,
  Nov. 2017.

\bibitem{rabiner1989}
L.~Rabiner, ``A tutorial on hidden {Markov} models and selected applications in
  speech recognition,'' \emph{Proc. IEEE}, vol.~77, no.~2, pp. 257--286, Feb.
  1989.

\bibitem{goodfellow2014}
I.~Goodfellow, J.~Pouget-Abadie, M.~Mirza, B.~Xu, D.~Warde-Farley, S.~Ozair,
  A.~Courville, and Y.~Bengio, ``Generative adversarial nets,'' in
  \emph{Advances Neural Inf. Process. Syst.}, vol.~27, 2014, pp. 1--9.

\bibitem{kingma2014}
D.~P. Kingma and M.~Welling, ``Auto-encoding variational {Bayes},'' in
  \emph{Proc. Int. Conf. Learn. Representations}, 2014, pp. 1--14.

\bibitem{croitoru2023}
F.-A. Croitoru, V.~Hondru, R.~T. Ionescu, and M.~Shah, ``Diffusion models in
  vision: A survey,'' \emph{{IEEE} Trans. Pattern Anal. Mach. Intell.},
  vol.~45, no.~9, pp. 10\,850--10\,869, Sep. 2023.

\bibitem{gunduz2023}
D.~Gündüz, Z.~Qin, I.~E. Aguerri, H.~S. Dhillon, Z.~Yang, A.~Yener, K.~K.
  Wong, and C.-B. Chae, ``Beyond transmitting bits: Context, semantics, and
  task-oriented communications,'' \emph{{IEEE} J. Sel. Areas Commun.}, vol.~41,
  no.~1, pp. 5--41, Jan. 2023.

\bibitem{kipnis2021}
A.~Kipnis, S.~Rini, and A.~J. Goldsmith, ``The rate-distortion risk in
  estimation from compressed data,'' \emph{{IEEE} Trans. Inf. Theory}, vol.~67,
  no.~5, pp. 2910--2924, May 2021.

\bibitem{liu2021}
J.~Liu, W.~Zhang, and H.~V. Poor, ``A rate-distortion framework for
  characterizing semantic information,'' in \emph{Proc. IEEE Int. Symp. Inf.
  Theory (ISIT)}, 2021, pp. 2894--2899.

\bibitem{liu2022}
J.~Liu, S.~Shao, W.~Zhang, and H.~V. Poor, ``An indirect rate-distortion
  characterization for semantic sources: General model and the case of
  {Gaussian} observation,'' \emph{{IEEE} Trans. Commun.}, vol.~70, no.~9, pp.
  5946--5959, Sep. 2022.

\bibitem{guo2022}
\BIBentryALTinterwordspacing
T.~Guo, Y.~Wang, J.~Han, H.~Wu, B.~Bai, and W.~Han, ``Semantic compression with
  side information: A rate-distortion perspective,'' Aug. 2022. [Online].
  Available: \url{https://arxiv.org/abs/2208.06094}
\BIBentrySTDinterwordspacing

\bibitem{wang2022}
Y.~Wang, T.~Guo, B.~Bai, and W.~Han, ``The estimation-compression separation in
  semantic communication systems,'' in \emph{Proc. IEEE Inf. Theory Workshop
  (ITW)}, Nov. 2022, pp. 315--320.

\bibitem{xiao2022}
Y.~Xiao, X.~Zhang, Y.~Li, G.~Shi, and T.~Başar, ``Rate-distortion theory for
  strategic semantic communication,'' in \emph{Proc. IEEE Inf. Theory Workshop
  (ITW)}, Nov. 2022, pp. 279--284.

\bibitem{stavrou2023}
P.~Stavrou and M.~Kountouris, ``The role of fidelity in goal-oriented semantic
  communication: A rate distortion approach,'' \emph{{IEEE} Trans. Commun.},
  vol.~71, no.~7, pp. 3918--3931, Jul. 2023.

\bibitem{wang1989}
S.~Wang and A.~Gersho, ``Phonetically-based vector excitation coding of speech
  at 3.6 kbps,'' in \emph{Proc. IEEE Int. Conf. Acoust., Speech and Signal
  Process. (ICASSP)}, May 1989, pp. 49--52 vol.1.

\bibitem{das1999}
A.~Das, A.~DeJaco, S.~Manjunath, A.~Ananthapadmanabhan, J.~Huang, and E.~Choy,
  ``Multimode variable bit rate speech coding: an efficient paradigm for
  high-quality low-rate representation of speech signal,'' in \emph{Proc. IEEE
  Int. Conf. Acoust., Speech, and Signal Process. (ICASSP)}, Mar. 1999, pp.
  2307--2310.

\bibitem{wang2023}
H.~Wang, Q.~Li, H.~Sun, Z.~Chen, Y.~Hao, J.~Peng, Z.~Yuan, J.~Fu, and Y.~Jiang,
  ``{VaBUS}: Edge-cloud real-time video analytics via background understanding
  and subtraction,'' \emph{{IEEE} J. Sel. Areas Commun.}, vol.~41, no.~1, pp.
  90--106, Jan. 2023.

\bibitem{wang2024}
R.~Wang, Y.~Hao, Y.~Miao, L.~Hu, and M.~Chen, ``{RT3C}: Real-time crowd
  counting in multi-scene video streams via cloud-edge-device collaboration,''
  \emph{{IEEE} Trans. Serv. Comput.}, vol.~17, no.~4, pp. 1739--1752, Jul.
  2024.

\bibitem{gray2001}
R.~Gray, ``{Gauss} mixture vector quantization,'' in \emph{Proc. IEEE Int.
  Conf. Acoust., Speech and Signal Process. (ICASSP)}, 2001, pp. 1769--1772.

\bibitem{subramaniam2003}
A.~Subramaniam and B.~Rao, ``{PDF} optimized parametric vector quantization of
  speech line spectral frequencies,'' \emph{{IEEE} Trans. Speech Audio
  Process.}, vol.~11, no.~2, pp. 130--142, Mar. 2003.

\bibitem{zhao2008}
D.~Zhao, J.~Samuelsson, and M.~Nilsson, ``On entropy-constrained vector
  quantization using {Gaussian} mixture models,'' \emph{{IEEE} Trans. Commun.},
  vol.~56, no.~12, pp. 2094--2104, Dec. 2008.

\bibitem{gray2023}
\BIBentryALTinterwordspacing
R.~Gray, \emph{Entropy and Information Theory}, 1st~ed.\hskip 1em plus 0.5em
  minus 0.4em\relax New York: Springer-Verlag, 1990, corrected version of June
  2023. [Online]. Available: \url{https://ee.stanford.edu/\~{}gray/it.html}
\BIBentrySTDinterwordspacing

\bibitem{arimoto1972}
S.~Arimoto, ``An algorithm for computing the capacity of arbitrary discrete
  memoryless channels,'' \emph{{IEEE} Trans. Inf. Theory}, vol.~18, no.~1, pp.
  14--20, Jan. 1972.

\bibitem{blahut1972}
R.~Blahut, ``Computation of channel capacity and rate-distortion functions,''
  \emph{{IEEE} Trans. Inf. Theory}, vol.~18, no.~4, pp. 460--473, Jul. 1972.

\bibitem{elias1975}
P.~Elias, ``Universal codeword sets and representations of the integers,''
  \emph{{IEEE} Trans. Inf. Theory}, vol.~21, no.~2, pp. 194--203, Mar. 1975.

\end{thebibliography}
\end{document}